\documentclass[twocolumn]{revtex4}

\usepackage{graphicx} 

\def\al{\alpha}
\def\be{\beta}
\def\ga{\gamma}
\def\de{\delta}

\def\ve{\varepsilon}
\def\ze{\zeta}
\def\et{\eta}

\def\ka{\kappa}
\def\la{\lambda}

\def\rh{\rho}

\def\om{\omega}

\def\De{\Delta}

\def\La{\Lambda}
\def\Si{\Sigma}

\def\Om{\Omega}

\def\vev#1{\langle {#1}\rangle}

\def\ket#1{|{#1}\rangle}
\def\fr#1#2{{{#1} \over {#2}}}
\def\frac#1#2{{\textstyle{{#1}\over {#2}}}}
\def\half{{\textstyle{1\over 2}}}
\def\lsim{\mathrel{\rlap{\lower4pt\hbox{\hskip1pt$\sim$}}
    \raise1pt\hbox{$<$}}}
\def\gsim{\mathrel{\rlap{\lower4pt\hbox{\hskip1pt$\sim$}}
    \raise1pt\hbox{$>$}}}
\def\sqr#1#2{{\vcenter{\vbox{\hrule height.#2pt
         \hbox{\vrule width.#2pt height#1pt \kern#1pt
         \vrule width.#2pt}
         \hrule height.#2pt}}}}

\def\pr#1{{#1}^\prime}

\def\abs#1{\left|{#1}\right|}

\def\cs{ ^{133}{\rm Cs} }
\def\rb{ ^{87}{\rm Rb} }

\def\tb{\tilde{b}} 
\def\td{\tilde{d}} 
\def\tg{\tilde{g}} 
\def\tc{\tilde{c}} 
\def\tH{\tilde{H}} 

\def\tmf{ \widetilde{m}_F }
\def\hmf{ \widehat{m}_F }

\def\bo{$\be_{\oplus}$} 
\def\bs{$\be_s$} 
\def\yep{1} 
\def\nop{-} 
\def\phb{\phantom{\Big[}}
\def\calT{{\cal T}} 

\def\etal{{\it et al.}}

\def\codt{\cos{\om_s T_s}}
\def\sodt{\sin{\om_s T_s}}
\def\ctodt{\cos{2\om_s T_s}}
\def\stodt{\sin{2\om_s T_s}} 
\def\som{\om^\sharp} 
\def\ca{c_\al} 
\def\sa{s_\al} 
\def\cz{c_\ze} 
\def\sz{s_\ze} 
\def\ce{c_\et} 
\def\se{s_\et} 
\def\cto{c_{\Om T}} 
\def\sto{s_{\Om T}} 
\def\cta{c_{2\al}} 
\def\sta{s_{2\al}} 
\def\ctz{c_{2\ze}} 
\def\stz{s_{2\ze}}

\def\citelabel#1{ \label{#1} } 
\def\labeleeq#1{ \label{#1} \eeq } 
\def\labeleea#1{ \label{#1} \eea }

\def\a{$a_\mu$}
\def\b{$b_\mu$}
\def\c{$c_{\mu\nu}$}
\def\d{$d_{\mu\nu}$}
\def\e{$e_\mu$}
\def\f{$f_\mu$}
\def\g{$g_{\la\mu\nu}$}
\def\H{$H_{\mu\nu}$}

\newcommand{\beq}{\begin{equation}}
\newcommand{\eeq}{\end{equation}}
\newcommand{\bea}{\begin{eqnarray}}
\newcommand{\eea}{\end{eqnarray}}
\newcommand{\rf}[1]{(\ref{#1})}

\newcommand{\Eq}[1]{Eq.\ (\ref{#1})}

\begin{document}

\title{Probing Lorentz and CPT violation with space-based experiments}

\author{Robert Bluhm,$^a$
V.\ Alan Kosteleck\'y,$^b$
Charles D.\ Lane,$^c$
and Neil Russell$^d$ }

\address{$^a$Physics Department, 
Colby College, Waterville, ME 04901\\
$^b$Physics Department, 
Indiana University, Bloomington, IN 47405\\
$^c$Physics Department, 
Berry College, Mount Berry, GA 30149\\
$^d$Physics Department, 
Northern Michigan University, Marquette, MI 49855\\
\rm \phantom{xxxxxxxxxx} (IUHET 455, June 2003)
\phantom{xxxxxxxxxx}}

\begin{abstract}
Space-based experiments offer sensitivity to numerous 
unmeasured effects involving Lorentz and CPT violation.
We provide a classification of clock sensitivities 
and present explicit expressions for time variations
arising in such experiments from nonzero
coefficients in the Lorentz- and CPT-violating Standard-Model Extension.

\end{abstract}
\maketitle

\section{Introduction}

Unification of the fundamental forces in nature
is expected to occur at the Planck scale,
$m_P\simeq 10^{19}$ GeV,
where quantum physics and gravity meet.
Performing experiments with energies 
at this scale is presently infeasible,
but suppressed signals might be detectable
in exceptionally sensitive tests. 
Searching for violations of relativity 
that might occur at the Planck scale
via the breaking of Lorentz and CPT symmetry 
is one promising approach
to uncovering Planck-scale physics 
\cite{cpt01}.

At low energies relative to the Planck scale,
observable effects of Lorentz violation 
are described by a general effective quantum field theory
constructed using the particle fields in the Standard Model.
This theory,
called the Standard-Model Extension (SME)
\cite{ck},
allows for general coordinate-independent
violations of Lorentz symmetry.
It provides a connection to the Planck scale
through operators of nonrenormalizable dimension 
\cite{kle}.
CPT violation implies Lorentz violation
\cite{owg},
so the SME also describes general effects from CPT violation.

Various origins are possible for the Lorentz and CPT violation
described by the SME.
An elegant and generic mechanism is 
spontaneous Lorentz violation,
originally proposed in the context of string theory
and field theories with gravity
\cite{ks}
and subsequently extended to include CPT violation
in string theory
\cite{kp}.
Noncommutative field theories offer 
another popular field-theoretic context for Lorentz violation,
in which realistic models form a subset of the SME
\cite{ncqed}.
Lorentz violation has also been proposed as a feature of certain 
non-string approaches to quantum gravity,
including loop quantum gravity
and related models of spacetime foam
\cite{qg},
the random dynamics approach
\cite{fn},
and multiverse models
\cite{bj}.

Various types of sensitive experiments can search
for the low-energy signals predicted by the SME.
In this work,
we consider clock-comparison experiments
with clocks co-located on a space platform,
which are known to offer a broad range of options
for Planck-sensitive tests of Lorentz and CPT symmetry
\cite{spaceexpt,km}.
Promising possibilities are offered by various
experiments planned for flight on 
the International Space Station (ISS),
including the 
ACES
\cite{aces},
PARCS
\cite{parcs},
RACE
\cite{race},
and SUMO
\cite{sumo}
missions.
The first three of these presently involve
atomic clocks with $\cs$, $\rb$
and a H maser
\cite{rv},
while the fourth uses a superconducting
microwave oscillator. 

Clock-comparison experiments in laboratories on the Earth
\cite{ccexpt,lh,db,dp,kla}
have already demonstrated exceptional sensitivity 
to spacetime anisotropies at the Planck scale.
These experiments monitor the frequency variations
of a Zeeman hyperfine transition
as the instantaneous atomic inertial frame 
changes orientation.
Typically,
a pair of clocks involving different atomic species 
and co-located in the laboratory
is compared as the Earth rotates.
Several other types of experiments
are also sensitive to Planck-scale effects
predicted by the SME,
including ones involving 
photons
\cite{km,photonexpt,photonth},
hadrons
\cite{hadronexpt,hadronth},
muons
\cite{muons},
and electrons
\cite{eexpt,eexpt2}.

In the present work,
we perform a general analysis
of clock-comparison experiments
involving atomic clocks on a satellite
such as the ISS.
To take advantage of the relatively high velocities
available in space,
we incorporate leading-order relativistic effects 
arising from clock boosts.
A framework for general calculations of this type 
is presented,
and detailed expressions 
that allow for satellite and Earth boosts
are derived for observables 
in a standard satellite mode.
Estimates are provided of the sensitivities
of experiments attainable on the ISS.

The paper is organized as follows. 
Section \ref{ClockInertialFrame} 
considers some aspects of the frequency shifts 
due to Lorentz violation 
that are experienced by a clock in a single inertial frame. 
In Sec.\ \ref{ClockNoninertialFrame}, 
we establish the link between a noninertial clock frame
on a space platform
and the standard Sun-based frame. 
Section \ref{Experiment} 
presents methods for extracting measurements of
coefficients for Lorentz violation
from experimental data
and estimates sensitivities for ISS-type missions.
We summarize in Sec.\ \ref{Summary}. 
Details of some calculations
are provided in some appendices.
Throughout this work,
we adopt the notation of Refs.\ \cite{ck,km}.

\section{Basics}
\citelabel{ClockInertialFrame}

Any Zeeman transition frequency $\om$ 
used to study Lorentz and CPT violation
can be written in the form
\beq
\om = f(B_3) + \de \om .
\labeleeq{FullFrequency} 
Here,
$B_3$ is the magnitude of the external magnetic field 
when projected along the quantization axis,
$f(B_3)$ is the transition frequency according to conventional physics,
and $\de\om$ contains all contributions from Lorentz and CPT violation.
All orientation dependence is contained in $B_3$ and $\de\om$;
in particular, 
the function $f$ has no orientation dependence except through $B_3$. 
Typically, 
$f$ depends on magnetic moments, 
angular-momentum quantum numbers, 
and similar quantities. 
For definiteness in what follows,
we suppose $f$ is invertible in a neighborhood
of the magnetic fields of interest
\cite{fn1},
and denote the inverse of $f$ by $f^{-1}(x)$. 
Also, 
we work at all orders in $B_3$
but neglect effects of size $o(B_3\de\om)$ 
and $o(\de\om^2)$,
which are known to be small.

For the transition 
$(F,m_F) \rightarrow (\pr{F},\pr{m}_F)$,
the frequency shift $\de\om$
can be written as
\bea
\de\om 
&=& \de E(F,m_F)- \de E(\pr{F},\pr{m}_F) ,
\labeleea{FrequencyComparison}  
where the atomic energy shifts $\de E (F,m_F)$
are induced by Lorentz and CPT violation.
These shifts can be calculated directly within the SME
using standard perturbation theory,
by obtaining the individual energy shifts 
for each constituent particle
and combining the results.
In the clock frame,
they are determined at leading order by a few
combinations of SME coefficients for Lorentz violation, 
conventionally denoted as
$\tb_3^w$, $\td_3^w$, $\tg_d^w$, $\tc_q^w$, $\tg_q^w$,
where the superscript $w$ is $p$ for the proton,
$n$ for the neutron, and $e$ for the electron.
These are the only quantities in the clock frame 
that can in principle be probed in clock-comparison experiments
with ordinary matter
\cite{kla}.

In the clock frame,
the atomic energy shift for state $\ket{F,m_F}$
can be written
\bea 
\de E(F,m_F) &=& 
\hmf \sum_w (\be_w\tb_3^w + \de_w\td_3^w + \ka_w\tg_d^w) 
\nonumber \\ 
&+& \tmf \sum_w (\ga_w\tc_q^w + \la_w\tg_q^w)
\quad .
\labeleea{AtomicShift}  
Here,
$\hmf$ and $\tmf$
are specific ratios of Clebsch-Gordan coefficients,
while 
$\be_w$, $\de_w$, $\ka_w$, $\ga_w$, $\la_w$
are specific expectation values 
of combinations of spin and momentum operators
in the extremal states $\ket{F, m_F=F}$.
For present purposes,
the details of these quantities are unnecessary;
they are given in Eqs.\ (7), (9), (10) 
of Ref.\ \cite{kla}.

Clock-comparison experiments 
typically involve two clocks
and corresponding transitions $A$, $B$
with frequency shifts
$\de\om_A$, $\de\om_B$,
located in an external magnetic field $B_3$.
The experimental signal of interest
is a modified difference between frequency shifts
of the form 
$\de\om_A - v \de\om_B$, 
where $v$ is an experiment-specific constant
related to the gyromagnetic ratios of the two clocks.
In typical arrangements,
$v$ is such that this signal vanishes 
in the absence of Lorentz violation. 
Note that the two transitions
may involve the same atomic species.

To bridge experiment and theory, 
it is useful to introduce a modified frequency difference 
$\som$ that represents the signal
for a large class of experimental situations
and offers a direct link to coefficients 
for Lorentz violation in the SME. 
For the two clock transitions $A$, $B$ 
with frequencies $\om_A$, $\om_B$
written in the form \rf{FullFrequency},
define 
$\som$ by 
\beq
\som := \om_A- f_A\left( f_B^{-1}(\om_B)\right) . 
\labeleeq{SharpOm1}  
By construction, 
$\som$ vanishes in the absence of Lorentz violation.
To the order in $\de\om$ at which we work,
this implies $\som$ is independent of the external magnetic field
even in the presence of Lorentz violation.
It is therefore reasonable to adopt 
this definition of $\som$ as the ideal observable
for Lorentz and CPT violation.
In what follows,
we first obtain a general theoretical expression for $\som$
and then consider some experimental issues.

We next show that $\som$ is determined theoretically
by the equation 
\beq 
\som = \de\om_A - v \de\om_B , 
\labeleeq{SharpOm2}  
where 
\beq 
v = \left( \fr {df_A}{dB_3}\bigg/\fr{df_B}{dB_3}\right)
    \bigg\vert_{B_3=0}  ,
\labeleeq{DefineV} 
and $\de\om_A$, $\de\om_B$
are given by Eq.\ \rf{FrequencyComparison}. 
This expression for $v$ is valid to all orders in $B_3$.
When combined with Eqs.\ \rf{FullFrequency},
\rf{FrequencyComparison}, and \rf{AtomicShift}, 
the above two equations allow calculation of $\som$.

To prove Eqs.\ \rf{SharpOm2} and \rf{DefineV},
we proceed as follows.
For each transition
of the form \rf{FullFrequency},
define an effective magnetic field
$B^{\rm eff} = f^{-1}(\om)$.
In the special case of no Lorentz violation,
$B^{\rm eff}$
is identical to the actual magnetic field $B_3$,
so the difference $B^{\rm eff}_A - B^{\rm eff}_B$ 
between the transitions $A$, $B$ is zero.
However, in general we have 
\begin{widetext}
\bea
B^{\rm eff}_A - B^{\rm eff}_B 
&=& f_A^{-1}(\om_A) - f_B^{-1}(\om_B)
 = f_A^{-1}\left[ f_A(B_3)+\de\om_A\right]
    -f_B^{-1}\left[ f_B(B_3)+\de\om_B\right]
\nonumber \\
 &=& \left.
\de\om_A
\fr{df_A^{-1}}{dx}\right|_{x=f_A(B_3)} 
    -\left.
\de\om_B 
\fr{df_B^{-1}}{dx}\right|_{x=f_B(B_3)} 
 + o(\de\om)^2 ,
\labeleea{DeriveSharpFreq1}  
where Taylor expansions in $\de\om_A$ 
and $\de\om_B$ have been performed. 
This implies 
\bea 
\om_A &=&
 f_A \left[ f_B^{-1}(\om_B) 
+\left.
\de\om_A
\fr{df_A^{-1}}{dx}\right|_{x=f_A(B_3)} 
 -\left.
\de\om_B  
\fr{df_B^{-1}}{dx}\right|_{x=f_B(B_3)} 
\right]
\nonumber \\
 &=& f_A\left( f_B^{-1}(\om_B) \right)
  + \left.\fr{df_A}{dy}\right|_{y=f_B^{-1}(\om_B)}
 \left[ \left.
\de\om_A
\fr{df_A^{-1}}{dx}\right|_{x=f_A(B_3)} 
 -\left.
\de\om_B 
\fr{df_B^{-1}}{dx}\right|_{x=f_B(B_3)} 
\right] ,
\labeleea{DeriveSharpFreq3}
\end{widetext} 
where another Taylor expansion has been performed.
Within factors of size $o(B_3\de\om)$,
we can set $B_3=0$ on the right-hand side of this equation, 
except for the term $f_A[f_B^{-1}(\om_B)]$.
Applying the identity
$({df^{-1}}/{dx})|_{x=f(B_3)} = ({df}/{dB_3})|_{B_3}^{-1}$
then yields Eqs.\ \rf{SharpOm2} and \rf{DefineV}.

As a first example of calculation with these results,
consider the special case of linear dependence on $B_3$.
Suppose for each transition we can write
$f(B_3)=c+\mu B_3$,
where $c$ and $\mu$ are constants
for each transition.
Then, we find
\beq
\som =
\om_A - \fr{\mu_A}{\mu_B}\om_B
- \left[ c_A -\fr{\mu_A}{\mu_B}c_B \right] .
\labeleeq{SharpFreqLinearB} 
In this case,
it suffices to study the combination
$\om_A-{\mu_A}\om_B/{\mu_B}$
and neglect the constants,
since clock-comparison experiments are only sensitive to 
orientation-dependent effects.

For a more complicated example,
consider the special case of a quadratic dependence on $B_3$.
Suppose for each transition we can write 
$ f(B) = c + \mu B + \rh B^2$,
where again $c$, $\mu$, $\rh$ are constants 
for each transition.
As always, $\som$ is relatively simple
when expressed in terms of frequency shifts for Lorentz violation: 
$\som = \de\om_A - {\mu_A}\de\om_B/{\mu_B}$.
However,
in terms of the individual frequencies $\som$ is
\bea
\som &=&
 \om_A - \fr{\rh_A}{\rh_B}\om_B 
\nonumber \\ 
 && - \left(\fr{\mu_A}{2\rh_B}-\fr{\mu_B\rh_A}{2\rh_B^2}\right)
        \sqrt{ \mu_B^2+4\rh_B(\om_B-c_B)}
\nonumber \\
 && + {\rm {~constant~terms}} .
\labeleea{SharpFreqQuadraticB2} 
Note that the previous linear example 
is a nontrivial limit of this one 
because $f^{-1}$ behaves badly as $\rh\to 0$.

A clock-comparison experiment to probe Lorentz violation
can proceed in several ways.
The most direct method is to
measure $\om_A$ and $\om_B$ at each instant. 
The results are then combined according to \Eq{SharpOm1}
to give an experimental value of $\som$,
which may be compared to 
the theoretical calculation in \Eq{SharpOm2}. 
A potentially significant disadvantage of this method 
is that achieving the desired sensitivity
requires exquisitely precise knowledge 
of the functions $f_A$ and $f_B$
and the parameters on which they depend. 

A different procedure can be adopted  
that requires no knowledge of the functions $f_A$ and $f_B$. 
Suppose $\om_B$ is forced to be constant, 
perhaps by applying a feedback magnetic field 
\cite{lh,db}.
Then, $f_A[f_B^{-1}(\om_B)]$ is constant,
so $\som=\om_A$
up to a constant irrelevant for experimental purposes.
Thus,
if $\om_B$ is held constant
and the transitions $A$ and $B$ involve clocks
subject to the same instantaneous magnetic field, 
it follows that $\om_A=\som=\de\om_A-v\de\om_B$.
Then,
$\om_A$ is sensitive purely to Lorentz-violating effects
and can be interpreted 
without detailed knowledge of $f_A$ and $f_B$.
This procedure may offer practical advantages
for experiments in environments with fluctuating magnetic fields
such as those anticipated for the ISS experiments.

\section{Frame transformations}
\citelabel{ClockNoninertialFrame} 

In a clock-comparison experiment,
the instantaneous clock frame is continuously changing
due to the orbital and rotational motion 
of the space-based laboratory
\cite{fn2}.
The quantities 
$\tb_3^w$, $\td_3^w$, $\tg_d^w$, $\tc_q^w$, $\tg_q^w$,
therefore vary in time,
with frequencies determined 
by the orbital and rotation periods of the laboratory.
This time variation can be obtained explicitly
by converting 
$\tb_3^w$, $\td_3^w$, $\tg_d^w$, $\tc_q^w$, $\tg_q^w$,
from the laboratory frame 
with coordinates $(0,1,2,3)$
to a specified nonrotating frame
with coordinates $(T,X,Y,Z)$. 

Following Refs.\ \cite{spaceexpt,km},
in this work we adopt for the standard nonrotating frame
a natural Sun-centered celestial equatorial frame.
This frame is approximately inertial over thousands of years.
It is therefore suitable 
for the study of leading-order boost effects
due to the Earth and satellite orbital motions.
The results of all clock-comparison experiments to date 
can be regarded as having been reported in this frame.

\begin{figure}
\scalebox{0.4}{ \includegraphics{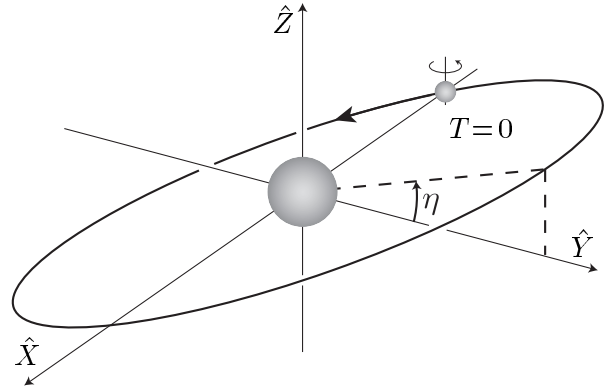} }
\caption{Orbit of Earth in Sun-based frame}
\citelabel{fig:EarthOrbit}
\end{figure}

In the Sun-based frame,
the spatial origin coincides with the center of the Sun.
The unit vector $\hat{Z}$ is parallel to the Earth's rotational axis, 
$\hat{X}$ points to the vernal equinox on the celestial sphere,
and $\hat{Y}$ completes the right-handed system.
The time $T$ is measured by a clock fixed at the origin,
with $T=0$ chosen as the vernal equinox in the year 2000.
Note that the vectors $\hat{X}$, $\hat{Y}$ lie in the 
Earth's equatorial plane,
which itself is at an angle of $\et\approx 23^\circ$ 
to the Earth's orbital plane.
Note also that the Earth is on the negative $X$ axis
at time $T=0$. 
See Fig.\ \ref{fig:EarthOrbit}. 

For a space-based experiment,
the time variation of the clock frequency is
determined by the satellite orbital and rotational motions.
To extract the leading-order effects
relevant for experiments on the Earth and on the ISS,
it suffices to approximate the orbits as circles.
Any ellipticity introduces time dependence 
at higher harmonics of orbital frequencies,
suppressed by even powers of the orbit eccentricity 
$\ve^2$.
For example,
a time dependence proportional to  
$\cos \om t$ under the circular-orbit approximation 
generates an order-$\ve^2$ dependence $\sim \ve^2\cos 3\om t$
for an elliptical orbit.
These harmonics appear only at subleading order
for any quantity that they modify.
For present purposes,
the circular approximation is reasonable because
$\ve_\oplus^2 \simeq 0.029$ for the Earth's orbit
and $\ve_{s}^2 \simeq 0.032$ for the ISS orbit.
However,
dedicated satellite missions could have strongly elliptical orbits,
in which case the higher harmonics would be of interest. 

Under the circular approximation,
the parameters of the Earth's orbit are
the mean orbital radius $R_\oplus$
and the mean orbital angular frequency 
$\Om_\oplus$.
The mean Earth orbital speed is 
$\be_\oplus=R_\oplus\Om_\oplus$.
The parameters for a circular satellite orbit 
around the Earth are taken as
the mean orbital radius
$r_s$, 
the mean satellite orbital angular frequency
$\om_s$, 
the angle $\ze$ between the Earth's rotation axis $\hat{Z}$ 
and the satellite orbital axis,
the azimuthal angle $\al$ 
between the satellite and the Earth orbital planes,
and a conveniently chosen reference time $T_0$
at which the satellite crosses the equatorial plane
on an ascending orbit.
See Figure \ref{fig:SatParams}.
It is also useful to introduce the satellite time 
measured in the Sun-based frame, $T_s=T-T_0$.
Note that the mean satellite speed 
with respect to the Earth's center is $\be_s=r_s\om_s$. 
For special limiting orbits,
$\om_s$ reduces to the usual sidereal frequency  
\cite{fn3}.
Note also that various perturbations typically cause $\al$ to precess.

\begin{figure}
\scalebox{0.5}{ \includegraphics{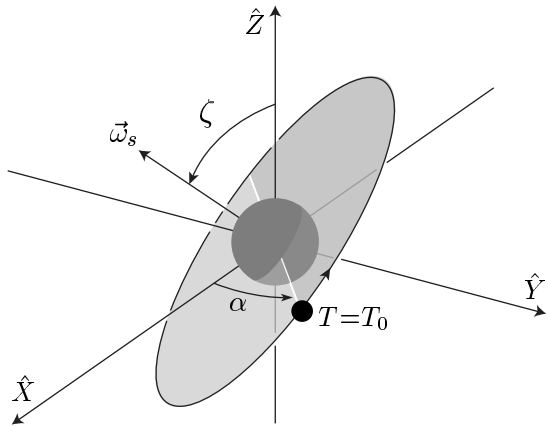} }
\caption{Parameters for definition of satellite orbit.
To simplify the presentation,
Earth is pictured as if it were translated
to the Sun-frame coordinate origin.}
\citelabel{fig:SatParams}
\end{figure}

\begin{table*} 
\setlength{\tabcolsep}{0.8mm} 
\renewcommand{\arraystretch}{0.9} 
\begin{tabular}{c|c|*7{c}*9{c}*8{c}} 
\hline \hline 
& $T_s$ & 
 $\tb_X$    &          & $\tb_T$     &          &          &          &          & $\td_{XY}$ &          & $\td_X$    & 
          &          &          & $\tg_-$    & $\tg_{XY}$ & $\tg_{ZX}$ & $\tg_{XZ}$ & $\tg_{DX}$ &          &          & 
 $\tc_X$    & $\tc_-$    & $\tc_{TX}$ &  
\\ 
& dep. & 
 $\tb_Y$    & $\tb_Z$    & $\tg_T$    & $\td_+$    & $\td_-$    & $\td_Q$    & $\tH_{JT}$ & $\td_{YZ}$ & $\td_{ZX}$ & $\td_Y$    & 
 $\td_Z$    & $\tg_c$    & $\tg_Q$    & $\tg_{TJ}$ & $\tg_{YX}$ & $\tg_{ZY}$ & $\tg_{YZ}$ & $\tg_{DY}$ & $\tg_{DZ}$ & $\tc_Q$    & 
 $\tc_Y$    & $\tc_Z$    & $\tc_{TY}$ & $\tc_{TZ}$ 
\\ 
\hline 
&&&&&&&&&& &&&&&&&&&& &&&&&
\\
$\tb_3$ & $\codt$  
 & \yep & \yep & \bo & \bo & \bo & \bo & \bo & \bo & \bo & \nop 
 & \nop & \bo & \nop & \nop & \nop & \nop & \nop & \nop & \nop & \nop 
 & \nop & \nop & \nop & \nop 
\\ 
 & $\sodt$ 
 & \yep & \nop & \bo & \bo & \bo & \bo & \bo & \bo & \nop & \nop 
 & \nop & \bo & \nop & \nop & \nop & \nop & \nop & \nop & \nop & \nop 
 & \nop & \nop & \nop & \nop 
\\ 
 & $\ctodt$ 
 & \nop & \nop & \bs & \nop & \bs & \bs & \nop & \bs & \bs & \nop 
 & \nop & \bs & \nop & \nop & \nop & \nop & \nop & \nop & \nop & \nop 
 & \nop & \nop & \nop & \nop 
\\ 
 & $\stodt$ 
 & \nop & \nop & \bs & \nop & \bs & \nop & \nop & \bs & \bs & \nop 
 & \nop & \bs & \nop & \nop & \nop & \nop & \nop & \nop & \nop & \nop 
 & \nop & \nop & \nop & \nop 
\\ 
 & const.  
 & \nop & \nop & \bs & \bs & \bs & \bs & \nop & \bs & \bs & \nop 
 & \nop & \bs & \nop & \nop & \nop & \nop & \nop & \nop & \nop & \nop 
 & \nop & \nop & \nop & \nop 
\\ 
&&&&&&&&&& &&&&&&&&&& &&&&&
\\
\hline 
&&&&&&&&&& &&&&&&&&&& &&&&&
\\
$\td_3$ & $\codt$ 
 & \nop & \nop & \bo & \bo & \bo & \bo & \bo & \bo & \bo & \yep 
 & \yep & \nop & \nop & \nop & \bo & \bo & \bo & \nop & \nop & \nop 
 & \nop & \nop & \nop & \nop 
\\ 
 & $\sodt$ 
 & \nop & \nop & \bo & \bo & \bo & \bo & \bo & \bo & \bo & \yep 
 & \nop & \nop & \nop & \nop & \bo & \bo & \bo & \nop & \nop & \nop 
 & \nop & \nop & \nop & \nop 
\\ 
 & $\ctodt$ 
 & \nop & \nop & \bs & \nop & \bs & \bs & \nop & \bs & \bs & \nop 
 & \nop & \nop & \nop & \nop & \bs & \bs & \bs & \nop & \nop & \nop 
 & \nop & \nop & \nop & \nop 
\\ 
 & $\stodt$ 
 & \nop & \nop & \nop & \nop & \bs & \nop & \nop & \bs & \bs & \nop 
 & \nop & \nop & \nop & \nop & \bs & \bs & \bs & \nop & \nop & \nop 
 & \nop & \nop & \nop & \nop 
\\ 
 & const.  
 & \nop & \nop & \bs & \bs & \bs & \bs & \nop & \bs & \bs & \nop 
 & \nop & \nop & \nop & \nop & \bs & \bs & \bs & \nop & \nop & \nop 
 & \nop & \nop & \nop & \nop 
\\ 
&&&&&&&&&& &&&&&&&&&& &&&&&
\\
\hline 
&&&&&&&&&& &&&&&&&&&& &&&&&
\\
$\tg_d$ & $\codt$ 
 & \nop & \nop & \bo & \nop & \nop & \nop & \nop & \nop & \nop & \nop 
 & \nop & \bo & \nop & \nop & \bo & \bo & \bo & \yep & \yep & \nop 
 & \nop & \nop & \nop & \nop 
\\ 
 & $\sodt$ 
 & \nop & \nop & \bo & \nop & \nop & \nop & \nop & \nop & \nop & \nop 
 & \nop & \bo & \nop & \nop & \nop & \bo & \bo & \yep & \nop & \nop 
 & \nop & \nop & \nop & \nop 
\\ 
 & $\ctodt$ 
 & \nop & \nop & \bs & \nop & \nop & \nop & \nop & \nop & \nop & \nop 
 & \nop & \bs & \nop & \nop & \bs & \bs & \bs & \nop & \nop & \nop 
 & \nop & \nop & \nop & \nop 
\\ 
 & $\stodt$ 
 & \nop & \nop & \bs & \nop & \nop & \nop & \nop & \nop & \nop & \nop 
 & \nop & \bs & \nop & \nop & \bs & \bs & \bs & \nop & \nop & \nop 
 & \nop & \nop & \nop & \nop 
\\ 
 & const.  
 & \nop & \nop & \bs & \nop & \nop & \nop & \nop & \nop & \nop & \nop 
 & \nop & \bs & \nop & \nop & \bs & \bs & \bs & \nop & \nop & \nop 
 & \nop & \nop & \nop & \nop 
\\ 
&&&&&&&&&& &&&&&&&&&& &&&&&
\\
\hline 
&&&&&&&&&& &&&&&&&&&& &&&&&
\\
$\tc_q$ & $\codt$ 
 & \nop & \nop & \nop & \nop & \nop & \nop & \nop & \nop & \nop & \nop 
 & \nop & \nop & \nop & \nop & \nop & \nop & \nop & \nop & \nop & \nop 
 & \nop & \nop & \bs & \bs 
\\ 
 & $\sodt$ 
 & \nop & \nop & \nop & \nop & \nop & \nop & \nop & \nop & \nop & \nop 
 & \nop & \nop & \nop & \nop & \nop & \nop & \nop & \nop & \nop & \nop 
 & \nop & \nop & \bs & \nop 
\\ 
 & $\ctodt$ 
 & \nop & \nop & \nop & \nop & \nop & \nop & \nop & \nop & \nop & \nop 
 & \nop & \nop & \nop & \nop & \nop & \nop & \nop & \nop & \nop & \yep 
 & \yep & \yep & \bo & \bo 
\\ 
 & $\stodt$ 
 & \nop & \nop & \nop & \nop & \nop & \nop & \nop & \nop & \nop & \nop 
 & \nop & \nop & \nop & \nop & \nop & \nop & \nop & \nop & \nop & \nop 
 & \yep & \yep & \bo & \bo 
\\ 
 & const.  
 & \nop & \nop & \nop & \nop & \nop & \nop & \nop & \nop & \nop & \nop 
 & \nop & \nop & \nop & \nop & \nop & \nop & \nop & \nop & \nop & \yep 
 & \yep & \yep & \bo & \bo 
\\ 
&&&&&&&&&& &&&&&&&&&& &&&&&
\\
\hline 
&&&&&&&&&& &&&&&&&&&& &&&&&
\\
$\tg_q$ & $\codt$ 
 & \nop & \nop & \nop & \nop & \nop & \nop & \nop & \nop & \nop & \nop 
 & \nop & \nop & \nop & \nop & \bs & \bs & \bs & \nop & \nop & \nop 
 & \nop & \nop & \nop & \nop 
\\ 
 & $\sodt$ 
 & \nop & \nop & \nop & \nop & \nop & \nop & \nop & \nop & \nop & \nop 
 & \nop & \nop & \nop & \nop & \bs & \nop & \bs & \nop & \nop & \nop 
 & \nop & \nop & \nop & \nop 
\\ 
 & $\ctodt$ 
 & \nop & \nop & \bo & \nop & \nop & \nop & \nop & \nop & \nop & \nop 
 & \nop & \bo & \yep & \yep & \bo & \bo & \bo & \nop & \nop & \nop 
 & \nop & \nop & \nop & \nop 
\\ 
 & $\stodt$ 
 & \nop & \nop & \bo & \nop & \nop & \nop & \nop & \nop & \nop & \nop 
 & \nop & \bo & \nop & \yep & \bo & \bo & \bo & \nop & \nop & \nop 
 & \nop & \nop & \nop & \nop 
\\ 
 & const.  
 & \nop & \nop & \bo & \nop & \nop & \nop & \nop & \nop & \nop & \nop 
 & \nop & \bo & \yep & \yep & \bo & \bo & \bo & \nop & \nop & \nop 
 & \nop & \nop & \nop & \nop 
\\ 
&&&&&&&&&& &&&&&&&&&& &&&&&
\\
\hline \hline 
\end{tabular} 
\caption{
Dependence of clock-frame coefficients on satellite time $T_s$ 
and on Sun-frame tilde coefficients. 
} 
\citelabel{TableTildes} 
\end{table*} 

The rotational motion of the satellite is specified
by giving its orientation as a function of time.
Two flight modes are commonly considered
\cite{ISS}, 
often denoted XVV and XPOP.
In XPOP mode,
the satellite orientation is fixed in the Sun-based frame
as it orbits the Earth.  
All clock signals from Lorentz violation
are due to boosts associated with 
the satellite orbital motion in this frame,
so they are suppressed by at least one power 
of $\be_\oplus$ or $\be_s$. 
In contrast,
for the XVV (``airplane'') mode,
the satellite rotates once in the Sun frame
each time it orbits the Earth, 
so its orientation is fixed relative to the instantaneous tangent 
to the satellite's circular orbit about the Earth.
Clock signals in this mode are due to both rotations and boosts, 
so they are sensitive to a wide variety of Lorentz-violating effects. 
In what follows,
we focus on the XVV mode.

In the space-based laboratory,
the coordinate system is defined as follows
\cite{spaceexpt,km}.
The 3 axis is taken along 
the satellite velocity with respect to the Earth. 
The 1 axis is chosen to point towards the center of Earth.
The 2 axis completes the right-handed system
and is oriented along the satellite orbital angular momentum
with respect to the Earth.
The clock orientation in the laboratory 
is typically determined by an applied magnetic field,
which establishes a quantization axis.
For definiteness,
we take the quantization axis as the 3 axis in this work. 
Other choices of quantization axis
can readily be calculated by our methods
\cite{km}.
Although the detailed time-varying signals are different,
no additional sensitivities to Lorentz violation
are obtained with other choices.

Combining information from the above 
frame, mode, and orientation choices
permits the construction 
of the explicit transformation $\calT$ 
between the Sun-based and laboratory frames.
Acting on vector components,
the transformation can be regarded as a matrix
with components ${\calT_\mu}^\Xi$
that depend on various velocities, frequencies, angles,
and Sun-frame times.
The derivation of this matrix
is provided in Appendix \ref{XformMatrix}. 
With this matrix,
the explicit time dependence of the quantities
$\tb_3^w$, $\td_3^w$, $\tg_d^w$, $\tc_q^w$, $\tg_q^w$,
can readily be calculated
in terms of the Sun-frame coefficients
$a_\Xi^w$, $b_\Xi^w$, $c_{\Xi\Pi}^w$, $d_{\Xi\Pi}^w$, 
$e_\Xi^w$, $f_\Xi^w$, $g_{\Xi\Pi\Si}^w$, $H_{\Xi\Pi}^w$
appearing in the fermion sector of the SME,
where $\Xi$, $\Pi$, $\Si$ are indices 
spanning the Sun-frame coordinates $(T,X,Y,Z)$.
For example, 
$b_3 = {\calT_3}^\Xi b_\Xi$,
and $d_{03} = {\calT_0}^\Xi {\calT_3}^\Pi d_{\Xi\Pi}$.

Due to the relatively involved spiral nature of the
satellite trajectory as observed in the Sun-frame,
the resulting explicit expressions 
for the quantities
$\tb_3^w$, $\td_3^w$, $\tg_d^w$, $\tc_q^w$, $\tg_q^w$,
are somewhat lengthy. 
It turns out to be simpler and natural
to express these in terms 
of certain special ``tilde'' combinations
of Sun-frame coefficients for Lorentz violation
\cite{fn4}.
These combinations are listed in  
Appendix \ref{DefineTildes}.
For each of the three species,
40 independent Sun-frame tilde coefficients
play a role at the level
of zeroth- and first-order relativistic effects
considered here.
There are therefore 120 linearly independent degrees of freedom
that can be probed in clock-comparison experiments 
with ordinary matter at this relativistic order.
Note that for each species the SME coefficients
\a, \b, \c, \d, \e, \f, \g, \H\
contain a total of 44 physically observable coefficients
at leading order in Lorentz violation
once unphysical field redefinitions have been fixed 
\cite{ck,cm,bek}, 
so four additional Sun-frame tilde coefficients are 
required to form a complete set of physical observables 
for clock-comparison experiments.
However,
these can appear at most as subleading-order relativistic effects
with signals suppressed by two powers of the velocities
$\be_\oplus$, $\be_s$
and are therefore not considered in this work.

The resulting expressions for 
$\tb_3^w$, $\td_3^w$, $\tg_d^w$, $\tc_q^w$, $\tg_q^w$,
are given in Appendix \ref{ExplicitTildes}. 
Each equation is a linear combination
of Sun-frame tilde coefficients.
The multiplicative factors are
constants of order 1,
sines or cosines of the angles $\al$, $2\al$, $\ze$, $2\ze$, $\et$,
and time oscillations involving sines or cosines of 
$\om_s T_s$, $2\om_s T_s$,
$\Om_\oplus T$. 
Note that
the terms involving $\Om_\oplus$ vary relatively slowly with time
because $\om_s \gg \Om_\oplus$.
The same is true of any precession time dependence 
in the orbital angle $\al$.
Note also that
the usual nonrelativistic dependence 
\cite{kla}
is recovered in the nonrelativistic limit
$\be_\oplus\rightarrow 0, \be_s\rightarrow 0$.

Insight into the content of these equations
can be obtained by separating each 
according to distinct satellite-frequency dependences
and classifying the resulting terms
according to velocity dependence. 
Since $\be_\oplus \simeq 10^{-4}$ for the Earth
and $\be_s \simeq 10^{-5}$ for the ISS,
the terms linear in the velocities 
are suppressed relative to the zeroth-order ones.
Table \ref{TableTildes} lists the decomposition 
of the equations in Appendix \ref{ExplicitTildes}
in accordance with this scheme. 
As an explicit example,
consider the variation of $\tc_q$
with the fundamental satellite frequency $\om_s$.
This is contained in the full expression 
for $\tc_q$ presented in Eq.\ \rf{Explicitcq},
from which the relevant terms can be extracted 
and rearranged in the form 
\bea
\tc_q &\supset&
\be_s ( 2s_\al c_\ze \tc_{TX} - 2c_\al c_\ze \tc_{TY}
  - 2s_\ze \tc_{TZ} ) \codt
 \nonumber \\
&& + \be_s ( 2c_\al \tc_{TX} + 2s_\al \tc_{TY} ) \sodt .
\labeleea{ExplicitcqShort} 
Table \ref{TableTildes}
separates the sine and cosine dependence of this expression
and lists factors of $\be_s$ 
in the appropriate columns
for the coefficients $\tc_{TX}$, $\tc_{TY}$, $\tc_{TZ}$.
In general,
this type of structural information about 
the equations in Appendix \ref{ExplicitTildes}
is useful in establishing sensitivities 
for different experiments.

\section{Signals and sensitivities}
\citelabel{Experiment} 

\begin{table*}
\begin{center}
\renewcommand{\arraystretch}{1.1}
\begin{tabular}{c*8{|c}*3{|c}}
\hline
\hline
 & $\cs$ & $\cs$ & $\cs$ & $\cs$ & $\cs$ 
& $\rb$ & $\rb$ & $\rb$ & $\rb$ & $\rb$ & $^{1}{\rm H}$  
\\ 
transition & $(4,0)$ & $(4,1)$ & $(4,-1)$ & $(4,0)$ & $(4,0)$ 
& $(2,1)$ & $(2,-1)$ & $(2,1)$ & $(2,0)$ & $(2,0)$ & $(1,1)$ \\ 
 & $\to(3,0)$ & $\to(3,1)$ & $\to(3,-1)$ & $\to(3,1)$ 
& $\to(3,-1)$ & $\to(1,1)$ & $\to(1,-1)$ 
 & $\to(1,0)$ & $\to(1,1)$ & $\to(1,-1)$ & $\to(1,0)$ \\ 
\hline 
$I$ & $7/2$ & $7/2$ & $7/2$ & $7/2$ & $7/2$ 
& $3/2$ & $3/2$ & $3/2$ & $3/2$ & $3/2$ & $1/2$ \\ 
$Z$ & $55$ & $55$ & $55$ & $55$ & $55$ & $37$ 
& $37$ & $37$ & $37$ & $37$ & $1$ \\ 
$N$ & $78$ & $78$ & $78$ & $78$ & $78$ & $50$ 
& $50$ & $50$ & $50$ & $50$ & $0$ \\ 
\hline 
Schmidt & $g_{7/2}$ & $g_{7/2}$ & $g_{7/2}$ 
& $g_{7/2}$ & $g_{7/2}$ & $p_{3/2}$ & $p_{3/2}$ 
 & $p_{3/2}$ & $p_{3/2}$ & $p_{3/2}$ & $s_{1/2}$ \\ 
nucleon & $$ & $$ & $$ & $$ & $$ & $$ & $$ & $$ & $$ & $$ & $$ \\ 
\hline 
$e^-$ state & $s_{1/2}$ & $s_{1/2}$ & $s_{1/2}$ 
& $s_{1/2}$ & $s_{1/2}$ & $s_{1/2}$ 
 & $s_{1/2}$ & $s_{1/2}$ & $s_{1/2}$ & $s_{1/2}$ & $s_{1/2}$ \\ 
\hline 
$\be_p$ & $[\fr{7}{9}]$ & $[\fr{7}{9}]$ & $[\fr{7}{9}]$ 
& $[\fr{7}{9}]$ & $[\fr{7}{9}]$ 
 & $[-1]$ & $[-1]$ & $[-1]$ & $[-1]$ & $[-1]$ & $-1$ \\ 
$\ga_p$ & $[-\fr{1}{9}K_p]$ & $[-\fr{1}{9}K_p]$ 
& $[-\fr{1}{9}K_p]$ & $[-\fr{1}{9}K_p]$ 
 & $[-\fr{1}{9}K_p]$ & $[-\fr{1}{15}K_p]$ 
& $[-\fr{1}{15}K_p]$ & $[-\fr{1}{15}K_p]$ 
 & $[-\fr{1}{15}K_p]$ & $[-\fr{1}{15}K_p]$ & $0$ \\ 
$\de_p$ & $[-\fr{7}{33}K_p]$ & $[-\fr{7}{33}K_p]$ 
& $[-\fr{7}{33}K_p]$ & $[-\fr{7}{33}K_p]$ 
 & $[-\fr{7}{33}K_p]$ & $[\fr{1}{5}K_p]$ 
& $[\fr{1}{5}K_p]$ & $[\fr{1}{5}K_p]$ 
 & $[\fr{1}{5}K_p]$ & $[\fr{1}{5}K_p]$ & $\fr{1}{3}K_p$ \\ 
$\ka_p$ & $[\fr{28}{99}K_p]$ & $[\fr{28}{99}K_p]$ 
& $[\fr{28}{99}K_p]$ & $[\fr{28}{99}K_p]$ 
 & $[\fr{28}{99}K_p]$ & $[-\fr{2}{5}K_p]$ 
& $[-\fr{2}{5}K_p]$ & $[-\fr{2}{5}K_p]$ 
 & $[-\fr{2}{5}K_p]$ & $[-\fr{2}{5}K_p]$ & $-\fr{1}{3}K_p$ \\ 
$\la_p$ & $[0]$ & $[0]$ & $[0]$ & $[0]$ & $[0]$ 
& $[0]$ & $[0]$ & $[0]$ & $[0]$ & $[0]$ & $0$ \\ 
\hline 
$\be_e$ & $-1$ & $-1$ & $-1$ & $-1$ & $-1$ & $-1$ 
& $-1$ & $-1$ & $-1$ & $-1$ & $-1$ \\ 
$\ga_e$ & $0$ & $0$ & $0$ & $0$ & $0$ & $0$ 
& $0$ & $0$ & $0$ & $0$ & $0$ \\ 
$\de_e$ & $\fr{1}{3}K_e$ & $\fr{1}{3}K_e$ & $\fr{1}{3}K_e$ 
& $\fr{1}{3}K_e$ & $\fr{1}{3}K_e$ 
 & $\fr{1}{3}K_e$ & $\fr{1}{3}K_e$ & $\fr{1}{3}K_e$ 
& $\fr{1}{3}K_e$ & $\fr{1}{3}K_e$ & $\fr{1}{3}K_e$ \\ 
$\ka_e$ & $-\fr{1}{3}K_e$ & $-\fr{1}{3}K_e$ 
& $-\fr{1}{3}K_e$ & $-\fr{1}{3}K_e$ & $-\fr{1}{3}K_e$ 
 & $-\fr{1}{3}K_e$ & $-\fr{1}{3}K_e$ & $-\fr{1}{3}K_e$ 
& $-\fr{1}{3}K_e$ & $-\fr{1}{3}K_e$ & $-\fr{1}{3}K_e$ \\ 
$\la_e$ & $0$ & $0$ & $0$ & $0$ & $0$ & $0$ 
& $0$ & $0$ & $0$ & $0$ & $0$ \\ 
\hline 
$s_1^p$ & $0$ & $-1/14$ & $1/14$ & $-9/28$ 
& $9/28$ & $-1/3$ & $1/3$ & $1/2$ & $-5/6$ & $5/6$ & $1$ \\ 
$s_2^p$ & $0$ & $-1/14$ & $-1/14$ & $-5/28$ 
& $-5/28$ & $-1$ & $-1$ & $1/2$ & $-3/2$ & $-3/2$ & $-$ \\ 
$s_1^e$ & $0$ & $1/2$ & $-1/2$ & $1/4$ 
& $-1/4$ & $1$ & $-1$ & $1/2$ & $1/2$ & $-1/2$ & $1$ \\ 
\hline \hline 
\end{tabular}
\renewcommand{\arraystretch}{1}
\caption{ Parameters for transition frequencies 
for experiments with Cs, Rb, and H clocks. 
}
\citelabel{SensCoeff}
\end{center}
\end{table*}

At this stage,
we can use the time dependence of the quantities
$\tb_3^w$, $\td_3^w$, $\tg_d^w$, $\tc_q^w$, $\tg_q^w$,
derived in the previous section 
to study the signals in clock-comparison experiments 
involving various atomic transitions.
We focus specifically on transitions 
$(F,m_F)\rightarrow (\pr{F},\pr{m}_F)$
in species scheduled for flight on the ISS: 
$\rb$, $\cs$, and H.
These have an even number of neutrons
and total electronic angular momentum $J=1/2$.
The generalization of our results 
to nuclei with an odd number of neutrons 
is straightforward.

The Lorentz-violating contribution $\de\om$ 
to the frequency of the transition
$(F,m_F) \rightarrow (\pr{F},\pr{m}_F)$
is given by
Eqs.\ \rf{FrequencyComparison} and \rf{AtomicShift}. 
With the above assumptions,
this frequency shift can be expressed as:
\bea 
\hskip -15pt
\de\om &=& 
s_1^p \left[ \be_p(l_j)\tb_3^p + \de_p(l_j)\td_3^p
  + \ka_p(l_j)\tg_d^p \right]
\nonumber \\ 
&& + s_2^p \left[ \ga_p(l_j)\tc_q^p
  + \la_p(l_j)\tg_q^p \right]
\nonumber \\
&& + s_1^e \left[ \be_e(0_{1/2})\tb_3^e + \de_e(0_{1/2})\td_3^e
  + \ka_e(0_{1/2})\tg_d^e \right] . 
\labeleea{GenlFreq} 
In this equation,
each $l_j$ refers to the Schmidt nucleon.
Except for the quantities $s_j^w$ outside the brackets,
all variables in Eq.\ \rf{GenlFreq}
are those appearing in Eq.\ \rf{AtomicShift}. 
The specific values of the quantities 
$\be_w$, $\ga_w$, $\de_w$, $\ka_w$, $\la_w$
are given as Eqs.\ (11) and (12) of Ref.\ \cite{kla}.
The values of $s_j^w$ depend on the transition,
and formul\ae\ for them are given below.
Note that similar equations valid 
for more general atoms would also involve
$s_2^e$, $s_1^n$, and $s_2^n$ terms.

The expressions for the $s_j^w$ 
can be classified according to the possible values of  
$F$ and $\pr{F}$.
There are four cases of interest.
For each case,
we give the expressions first 
in terms of combinations of Clebsch-Gordan coefficients
and angular-momentum quantum numbers,
and then directly in terms of $m_F$ and $\pr{m}_F$.
In all cases,
we define
$\De m_F := m_F - \pr{m}_F$
and 
$\De m_F^2 := m_F^2 - (\pr{m}_F)^2$.

The first case has
$F=\pr{F}=I+\half$,
for which we obtain
\bea
s_1^p &= \hmf - \pr{\hmf}
&= \left[\frac{2}{2I+1}\right] \De m_F ,
\nonumber \\
s_2^p &= \tmf - \pr{\tmf}
&= \left[\frac{3}{I(2I+1)}\right] \De m_F^2 ,
\nonumber \\
s_1^e &= \hmf - \pr{\hmf}
&= \left[\frac{2}{2I+1}\right] \De m_F .
\labeleea{sCase1}
The second case has
$F= \pr{F}=I-\half$, 
and we find
\bea
s_1^p =& (\hmf - \pr{\hmf}) \frac{(I+1)(2I-1)}{I(2I+1)}
&= \left[\frac{2I+2}{I(2I+1)}\right] \De m_F ,
\nonumber \\
s_2^p =& (\tmf - \pr{\tmf}) \frac{(I-1)(2I+3)}{I(2I+1)}
&= \left[\frac{3(2I+3)}{I(2I+1)(2I-1)}\right] \De m_F^2 ,
\nonumber \\
s_1^e =& (\hmf - \pr{\hmf}) \frac{1-2I}{1+2I}
&= \left[\frac{-2}{2I+1}\right] \De m_F .
\labeleea{sCase2} 
The third case has
$F=I+\half, \pr{F}=I-\half$,
giving
\bea
s_1^p &=& \hmf - \frac{(I+1)(2I-1)}{I(2I+1)} \pr{\hmf} 
\nonumber \\ 
&=& \left[\frac{2}{2I+1}\right]\De m_F
- \left[\frac{2}{I(2I+1)}\right]\pr{m}_F ,
\nonumber \\
s_2^p &=& \tmf - \frac{(I-1)(2I+3)}{I(2I+1)} \pr{\tmf} 
\nonumber \\ 
&=& \left[\frac{3}{I(2I+1)}\right]\De m_F^2
- \left[\frac{12}{I(2I+1)(2I-1)}\right] (\pr{m}_F)^2 ,
\nonumber \\
s_1^e &=& \hmf - \frac{1-2I}{1+2I} \pr{\hmf} 
\nonumber \\ 
&=& \left[\frac{2}{2I+1}\right] \De m_F
+ \left[\frac{4}{2I+1}\right] \pr{m}_F .
\labeleea{sCase3} 
The final case has
$F=I-\half, \pr{F}=I+\half$,
for which
\bea
s_1^p &=& \frac{(I+1)(2I-1)}{I(2I+1)} \hmf - \pr{\hmf} 
\nonumber \\ 
&=& \left[\frac{2}{2I+1}\right] \De m_F
+ \left[\frac{2}{I(2I+1)}\right] m_F ,
\nonumber \\
s_2^p &=& \frac{(I-1)(2I+3)}{I(2I+1)} \tmf - \pr{\tmf}
\nonumber \\ 
&=& \left[\frac{3}{I(2I+1)}\right] \De m_F^2
+ \left[\frac{12}{I(2I+1)(2I-1)}\right] m_F^2 ,
\nonumber \\
s_1^e &=& \frac{1-2I}{1+2I} \hmf - \pr{\hmf}
\nonumber \\ 
&=& \left[\frac{2}{2I+1}\right] \De m_F
- \left[\frac{4}{2I+1}\right] m_F .
\labeleea{sCase4}
 
Various results can be obtained from these expressions
and Eq.\ \rf{GenlFreq}.
For example,
it follows directly that a nonzero signal occurs 
for all $\De F=\pm 1$, $\De m_F=0$
transitions except for the special case 
$m_F=\pr{m}_F=0$. 
This demonstrates that the standard clock transitions 
are insensitive to Lorentz-violating effects,
in agreement with previous results
\cite{kla}.
Useful special cases of immediate relevance 
to experiments on the ISS can also be extracted.
Thus,
for a $\cs$ clock with 
$I=\frac{7}{2}$, $F=4\rightarrow 3$,
we find:
\bea
s_1^p &=& \frac{1}{4}\De m_F -\frac{1}{14} \pr{m}_F , 
\nonumber \\
s_2^p &=& \frac{3}{28}\De m_F^2 - \frac{1}{14} (\pr{m}_F)^2 ,
\nonumber \\
s_1^e &=& \frac{1}{4}\De m_F + \frac{1}{2} \pr{m}_F .
\labeleea{sCaseCs} 
Similarly, 
for a $\rb$ clock with
$I=\frac{3}{2}$, $F=2\rightarrow 1$,
we obtain:
\bea
s_1^p &=& \frac{1}{2}\De m_F -\frac{1}{3} \pr{m}_F ,
\nonumber \\
s_2^p &=& \frac{1}{2}\De m_F^2 - (\pr{m}_F)^2 ,
\nonumber \\
s_1^e &=& \frac{1}{2}\De m_F + \pr{m}_F .
\labeleea{sCaseRb}
For an H clock or maser,
$s_1^e = s_1^p = 1$. 
However,
$s_2^p$ is irrelevant:
the proton is in an $I=1/2$ state,
so there is no quadrupole effect
and the quantities $\ga_p$ and $\la_p$ vanish.

Table \ref{SensCoeff} summarizes some useful results
for species scheduled for flight on the ISS.
The first few rows of this table identify the transition
and list various properties of the species involved.
The nuclear spin is $I$, the proton number is $Z$,
and the neutron number is $N$.
The following entry fixes
the proton determining the ground-state properties
of the nucleus following the nuclear Schmidt model
\cite{sch},
together with its associated 
orbital and total angular momentum.
The electronic configuration is also given.
Ten rows list the relevant parameters
$\be_w$, $\de_w$, $\ka_w$, $\ga_w$, $\la_w$,
with values in brackets obtained 
under the assumptions of the Schmidt model. 
We define $K_p=\vev{p^2}/m_p^2$, 
which can be regarded as twice the kinetic energy per mass 
of the Schmidt-model proton, 
and define $K_e$ similarly for the valence electron. 
An estimate gives $K_p\simeq 10^{-2}$ 
for all species except $^1$H, 
for which $K_p\simeq 10^{-11}$,
and $K_e\simeq 10^{-5}$.
Finally, 
the numerical values of the $s_j^w$ are listed. 
Where these are nonzero,
a clock-comparison experiment with the specified transition
is sensitive to Lorentz violation.

At this stage,
enough information is at hand to extract estimated
experimental sensitivities. 
Suppose the results of an experiment 
measuring the modified frequency difference 
$\som$ of Eq.\ \rf{SharpOm1}
are fitted to the form
\bea
\som &=& \mbox{constant} + 2\pi\ve_{1,X}\codt + 2\pi\ve_{1,Y}\sodt
\nonumber \\
&& + 2\pi\ve_{2,X}\ctodt + 2\pi\ve_{2,Y}\stodt .
\labeleea{SharpFreqTime} 
Nonzero values of any of the $\ve_{a,J}$ indicate Lorentz violation.
Denote by $\ve_a$ the minimum of $\{ |\ve_{a,X}|, |\ve_{a,Y}| \}$.
Then, 
combining the theoretical analysis above
yields the following predicted dependence 
of $\ve_a$ on coefficients for Lorentz violation,
atomic and nuclear parameters, and geometrical factors:
\begin{widetext} 
\bea
2\pi\ve_{1} &=& \bigg|\sum_w\bigg[
   (s_1^{wA}\be_w^A - vs_1^{wB}\be_w^B) (\tb_J^w)
+ \be_\oplus \big[s_1^{wA}
(\be_w^A+\de_w^A+\ka_w^A)-vs_1^{wB}(\be_w^B+\de_w^B+\ka_w^B)\big]
      (\tb_T^w,\tg_T^w)
\nonumber \\
&& 
+ \be_\oplus \big[s_1^{wA}
(\be_w^A+\de_w^A)-vs_1^{wB}(\be_w^B+\de_w^B)\big]
    (\td_\pm^w, \td_Q^w, \td_{JK}^w, \tH_{JT}^w )
+ (s_1^{wA}\de_w^A - vs_1^{wB}\de_w^B) (\td_J^w)
\nonumber \\
&& 
+ (s_1^{wA}\ka_w^A - vs_1^{wB}\ka_w^B) (\tg_{DJ}^w)
+ \big[ \be_\oplus s_1^{wA}
(\de_w^A+\ka_w^A)-\be_\oplus vs_1^{wB}(\de_w^B+\ka_w^B)
           +\be_s(s_2^{wA}\la_w^A-vs_2^{wB}\la_w^B) \big]
    (\tg_{JK}^w)
\nonumber \\
&& 
+ \be_\oplus \big[s_1^{wA}
(\be_w^A+\ka_w^A)-vs_1^{wB}(\be_w^B+\ka_w^B)\big]
    (\tg_c^w)
+ \be_s (s_2^{wA}\ga_w^A - vs_2^{wB}\ga_w^B) (\tc_{TJ}^w)
\bigg]\bigg| ,
\label{inter}
\\
2\pi\ve_{2} &=& \bigg|\sum_w\bigg[
     \big[ \be_s s_1^{wA}
(\be_w^A+\de_w^A+\ka_w^A)-\be_s vs_1^{wB}(\be_w^B+\de_w^B+\ka_w^B)
           +\be_\oplus (s_2^{wA}\la_w^A-vs_2^{wB}\la_w^B) \big]
    (\tb_T^w,\tg_T^w)
\nonumber \\
&& 
+ \be_s \big[s_1^{wA}
(\be_w^A+\de_w^A)-vs_1^{wB}(\be_w^B+\de_w^B)\big]
    (\td_-^w, \td_Q^w, \td_{JK}^w)
+ \be_s (s_1^{wA}\be_w^A - vs_1^{wB}\be_w^B) (\tH_{JT}^w)
\nonumber \\
&&
+ \big[ \be_s s_1^{wA}
(\de_w^A+\ka_w^A)-\be_s vs_1^{wB}(\de_w^B+\ka_w^B)
           +\be_\oplus (s_2^{wA}\la_w^A-vs_2^{wB}\la_w^B) \big]
    (\tg_{JK}^w)
\nonumber \\
&&
+ \big[ \be_s s_1^{wA}
(\be_w^A+\ka_w^A)-\be_s vs_1^{wB}(\be_w^B+\ka_w^B)
           +\be_\oplus (s_2^{wA}\la_w^A-vs_2^{wB}\la_w^B) \big]
    (\tg_c^w)
+ \be_\oplus (s_2^{wA}\ga_w^A - vs_2^{wB}\ga_w^B)
    (\tc_{TJ}^w)
\nonumber \\
&&
+ (s_2^{wA}\ga_w^A - vs_2^{wB}\ga_w^B)
    (\tc_-^w,\tc_Q^w,\tc_J^w)
+ (s_2^{wA}\la_w^A - vs_2^{wB}\la_w^B)
    (\tg_-^w,\tg_Q^w,\tg_{TJ}^w)
\bigg]\bigg| .
\labeleea{Epsilons}
\end{widetext} 
In these equations,
superscripts $A$ and $B$ 
indicate quantities evaluated for transitions
$A$ and $B$, respectively, 
and $J$ takes the values $(X,Y,Z)$.
The coefficients for Lorentz violation 
enter as somewhat lengthy linear combinations 
of the type appearing in 
the equations of Appendix \ref{ExplicitTildes}.
These explicit combinations are omitted here for brevity,
being replaced instead with 
parentheses containing only the specific 
coefficients for Lorentz violation involved.

The above form of the equations is useful 
despite the brevity because
it allows relatively straightforward consideration
of sensitivities to the 
Sun-frame tilde coefficients for Lorentz violation.
We adopt here the strategy of Ref.\ \cite{kla},
in which numerical sensitivities
are obtained within the Schmidt model
under the plausible assumption that no substantial cancellations
occur among contributions from different 
Sun-frame tilde coefficients for Lorentz violation.
For example,
if $\ve_1$ is an experimental 
sensitivity to the time variation of $\som$,
then Eq.\ \rf{inter} implies 
the experiment has sensitivity
to each $|\tb_J^w |$ of
$\sim 2\pi\ve_1 (s_1^{wA}\be_w^A - vs_1^{wB}\be_w^B)^{-1}$.
Similarly,
the sensitivity to $|\tb_T^w |$ is 
$\sim 2\pi\ve_1 \be_\oplus^{-1} [s_1^{wA}(\be_w^A+\de_w^A+\ka_w^A) 
-vs_1^{wB}(\be_w^B+\de_w^B+\ka_w^B)]^{-1}$,
and so on. 
To obtain crude order of magnitude numerical estimates,
it suffices to approximate 
$\be_s\sim 10^{-5}$, 
$\be_\oplus\sim 10^{-4}$,
and to estimate nonzero values of the other parameters as follows: 
$\be_w\sim 1$, $s_a^w\sim 1$ for all species;
$\de,\ka,\ga,\la\sim 10^{-2}$ for protons
except for $^1$H where the nonzero values are only 
$\de,\ka\sim 10^{-11}$ for the proton;
and $\de,\ka,\ga,\la\sim 10^{-5}$ for electrons.
Sensitivity estimates of this type are reasonable
provided the various angles 
$\al$, $2\al$, $\ze$, $2\ze$, $\et$, $\Om_\oplus T$
lie away from multiples of $\pi/4$.
The orientation of the quantization axis within the satellite 
then makes little difference to the sensitivity.
However,
for any angles near a multiple of $\pi/4$,
sensitivity to one or more of the Sun-frame tilde 
coefficients can be lost.

\begin{table}
\begin{center}
\begin{tabular}{cccc} 
\hline \hline 
Coefficient & Proton & Neutron & Electron \\
\hline
$\tb_X$, $\tb_Y$        & -27[-27]& [-31]   & -27[-29]  \\
$\tb_Z$                 & -27     & -       & -27[-28]   \\
$\tb_T$                 & -23     & -       & -23     \\
$\tg_T$                 & -23     & -       & -23     \\
$\tH_{JT}$              & -23     & -       & -23     \\
$\td_\pm$               & -23     & -       & -23     \\
$\td_Q$                 & -23     & -       & -23     \\
$\td_{JK}$              & -23     & -       & -23     \\
$\td_X$, $\td_Y$        & -25[-25]& [-29]   & -22[-22]  \\
$\td_Z$                 & -25     & -       & -22     \\
$\tg_{DX}$,$\tg_{DY}$   & -25[-25]& [-29]   & -22[-22]  \\
$\tg_{DZ}$              & -25     & -       & -22     \\
$\tg_{JK}$              & -21     & -       & -18     \\
$\tg_c$                 & -23     & -       & -23     \\
$\tc_{TJ}$              & -20     & -       & -       \\
$\tc_-$                 & -25     & [-27]     & -       \\
$\tc_Q$                 & -25     & -       & -       \\
$\tc_X$, $\tc_Y$        & -25     & [-25]     & -       \\
$\tc_Z$                 & -25     & [-27]     & -       \\
$\tc_{TJ}$              & -21     & -       & -       \\
$\tg_-$                 & $\star$[$\star$] & [$\star$] & -       \\
$\tg_Q$                 & $\star$ & -       & -       \\
$\tg_{TX}$, $\tg_{TY}$  & $\star$[$\star$] & [$\star$] & -       \\
$\tg_{TZ}$              & $\star$[$\star$] & [$\star$] & -       \\ 
\hline \hline 
\end{tabular}
\caption{ Estimated sensitivity 
to coefficients for Lorentz violation
for ISS experiments with $\cs$ and $\rb$ clocks.
Existing bounds 
\cite{ccexpt,lh,db,dp,eexpt2}
are shown in brackets.}
\citelabel{PropSensTable}
\end{center}
\end{table}

Table \ref{PropSensTable} lists estimated sensitivities
to Sun-frame tilde coefficients for Lorentz violation
that might be attained in the planned
space-based clock-comparison experiments
with $\cs$ and $\rb$ clocks.
The base-10 logarithm of the sensitivity per GeV
is shown for each coefficient for Lorentz violation
and for each particle species.
For definiteness,
the clock sensitivity has been taken as 
$\ve_{1,2} 
\sim$ 50 $\mu$Hz,
which is comparable to that attained 
in a ground-based experiment with $\cs$ 
\cite{lh},
but the results shown are readily scaled for other values 
of $\ve_{1,2}$. 
A star in the table indicates a combination 
for which there is no sensitivity 
according to the nuclear Schmidt model
but probable sensitivity in a more realistic nuclear model.
A value in brackets indicates 
an existing bound from an Earth-based experiment
\cite{ccexpt,lh,db,dp,eexpt2}.
Given the approximations described above,
some caution in interpretation of the details of this table 
is advisable.
Nonetheless,
the table provides a measure of the broad scope 
of space-based tests of Lorentz symmetry,
and it shows that Planck-scale sensitivity 
for a wide spectrum of relativity tests is attainable.

Space-based experiments exploring the photon sector of the SME 
have been studied elsewhere 
\cite{km},
including the SUMO experiment
with superconducting microwave-cavity oscillators
that is presently scheduled for flight.
Experiments of this type could be profitably combined with 
the clock-comparison experiments discussed in the present work.
For example,
at the time of writing PARCS and SUMO 
are planned for simultaneous flight.
Several configurations of interest could then be considered,
including operating PARCS on a Lorentz-insensitive line
as a reference for relativity tests with SUMO,
or seeking $\om_s$ and $2\om_s$ signals
in a configuration with both the atomic clock and the cavity
operating in modes sensitive to Lorentz violation.

\section{Summary}
\citelabel{Summary} 

This work has studied clock-comparison experiments
on a space-based platform,
with specific emphasis placed
on forthcoming experiments 
on the International Space Station.
The theoretical framework adopted  
is the Standard-Model Extension, 
which describes general Lorentz and CPT violation.
The analysis yields predictions for signals 
at the ISS orbital and double-orbital frequencies,
along with slower variations associated with 
the Earth orbital motion.

The formalism we have presented 
applies to any space-based experiment 
with atomic clocks
and incorporates relativistic effects
at first boost order.
We have derived explicit expressions
for the observable effects in the special cases
of $\cs$, $\rb$, and H clocks on the ISS,
which are currently planned for flight
in the PARCS, ACES, and RACE missions.
These results,
which involve the fermion sector of the SME,
complement the photon-sector analysis 
of Lorentz-violation sensitivity performed
for the planned SUMO experiment with microwaves on the ISS.

We have obtained estimates for the attainable sensitivities
with these atomic-clock missions,
listed in Table \ref{PropSensTable}.
Numerous currently unmeasured coefficients for Lorentz violation 
could be studied in these experiments.
The results demonstrate that experiments of this type
offer potential sensitivity to
violations of relativity with Planck-scale reach.

\section*{ACKNOWLEDGMENTS}

This work was supported in part by the
National Aeronautics and Space Administration
under grant NAG8-1770,
by the United States Department of Energy
under grant DE-FG02-91ER40661,
and by the National Science Foundation
under grant PHY-0097982.

\appendix

\section{Transformation from Sun-based to satellite frame} 
\citelabel{XformMatrix}

In this appendix,
we derive the transformation matrix ${\calT_\mu}^\Xi$
introduced in Sec.\ \ref{ClockNoninertialFrame} 
that maps Sun-frame quantities to laboratory-frame ones.
The transformation $\calT$
can be expressed as the composition of
a boost $\La$ from the Sun frame
to the (nonrotating) rest frame of the center of the satellite,
followed by a rotation ${\cal R}$ 
from the (nonrotating) rest frame of the center of the satellite
to the (rotating) lab frame.
Each constituent transformation depends on time,
and each is understood to be instantaneous.
We first obtain an expression 
for the satellite position in the Sun frame,
then use this to derive the instantaneous satellite velocity 
in the Sun frame,
and finally combine information to construct 
the desired overall transformation.
The conventions we adopt are those 
given in Refs.\ \cite{spaceexpt,km}.

The position 
$\vec{X}_\oplus$
of the center of the Earth 
in the Sun-based frame is given by
\beq
\vec{X}_\oplus=
\left(
 \begin{array}{c}
  -R_\oplus\cos{\Om_\oplus T} \\
  -R_\oplus\cos{\et}\sin{\Om_\oplus T} \\
  -R_\oplus\sin{\et}\sin{\Om_\oplus T}
 \end{array}
\right) .
\labeleeq{EarthPosition}
The satellite position $\vec{X}_s$
in this frame is obtained by adding the position
of the satellite with respect to the Earth,
which gives
\begin{widetext} 
\beq
\vec{X}_s=
\left(
 \begin{array}{c}
  -R_\oplus\cos{\Om_\oplus T}
   +r_s\cos{\al}\cos{\om_s T_s}
   -r_s\cos{\ze}\sin{\al}\sin{\om_s T_s} \\
  -R_\oplus\cos{\et}\sin{\Om_\oplus T}
   +r_s\sin{\al}\cos{\om_s T_s}
   +r_s\cos{\al}\cos{\ze}\sin{\om_s T_s} \\
  -R_\oplus\sin{\et}\sin{\Om_\oplus T}
   +r_s\sin{\ze}\sin{\om_s T_s}
 \end{array}
\right) .
\labeleeq{SatPosition}

\end{widetext} 

Disregarding rotations for the moment,
the boost $\La$ from the Sun-based frame 
to the nonrotating instantaneous satellite rest frame 
is determined by 
the velocity $\vec{V}=d\vec{X_s}/dT$.
To lowest order in $|\vec{V}|$,
this is
\beq
\La = \left(
\begin{array}{c|ccc}
 1 & V_X & V_Y & V_Z \\
 \hline
 V_X & 1 & 0 & 0 \\
 V_Z & 0 & 1 & 0 \\
 V_Z & 0 & 0 & 1
\end{array}
\right) .
\labeleeq{BoostGenl} 
The required rotation ${\cal R}$ 
from the nonrotating instantaneous satellite rest frame 
to the laboratory frame on the satellite 
may be calculated using the velocity and acceleration
of the satellite with respect to the Earth,
$d(\vec{X}_s-\vec{X}_\oplus)/dT \sim \hat{z}$
and $d^2(\vec{X}_s-\vec{X}_\oplus)/dT^2 \sim \hat{x}$,
and the requirement that
\beq
\left(
\begin{array}{c}
 \hat{x} \\ \hat{y} \\ \hat{z}
\end{array}
\right) =
{\cal R} \left(
\begin{array}{ccc}
 \hat{X} \\ \hat{Y} \\ \hat{Z}
\end{array}
\right) .
\labeleeq{RRequirement} 

Given $\La$ and ${\cal R}$,
the overall instantaneous transformation ${\cal T}$
from the Sun-based frame to the laboratory frame
can be found:
\beq
{\cal T} =
\left(
\begin{array}{c|ccc}
 1 & 0 & 0 & 0 \\
\hline
 0 & & & \\
 0 & & {\cal R} & \\
 0 & & &
\end{array}
\right)
\La
\quad .
\labeleeq{FullTransform} 
With the approximations of Sec.\ \ref{ClockNoninertialFrame},
the components ${\calT_\mu}^\Xi$
of the transformation matrix (\ref{FullTransform}) 
are found to be
\bea
{\calT_0}^T &=& 1 ,
 \nonumber \\
{\calT_0}^X &=& \be_\oplus \sto
 - \be_s \sa \cz \codt
 - \be_s \ca \sodt ,
 \nonumber \\
{\calT_0}^Y &=& -\be_\oplus \ce \cto
 + \be_s \ca \cz \codt
 - \be_s \sa \sodt ,
 \nonumber \\
{\calT_0}^Z &=& -\be_\oplus \se \cto
 + \be_s \sz \codt ,
 \nonumber \\
{\calT_1}^T &=&  (\sa \ce \cto
 - \ca \sto )\be_\oplus\codt
 \nonumber \\ 
 &&  
\hskip-10pt
+(\ca \cz \ce \cto
 + \sz \se \cto
 + \sa \cz \sto )\be_\oplus\sodt ,
 \nonumber \\
{\calT_1}^X &=& -\ca \codt + \sa \cz \sodt ,
 \nonumber \\
{\calT_1}^Y &=& - \sa \codt - \ca \cz \sodt ,
 \nonumber \\
{\calT_1}^Z &=& -\sz \sodt ,
 \nonumber \\
{\calT_2}^T &=& \be_\oplus \ca \sz \ce \cto
 - \be_\oplus \cz \se \cto
 + \be_\oplus \sa \sz \sto ,
 \nonumber \\
{\calT_2}^X &=& \sa \sz ,
 \nonumber \\
{\calT_2}^Y &=& -\ca \sz ,
 \nonumber \\
{\calT_2}^Z &=& \cz ,
 \nonumber \\
{\calT_3}^T &=& \be_s 
 + (\sa \ce \cto
 - \ca \sto )\be_\oplus \sodt
 \nonumber \\ 
 &&
\hskip-10pt
+ (- \ca \ce \cz \cto
 - \sz \se \cto
 - \sa \cz \sto )\be_\oplus\codt ,
 \nonumber \\ 
{\calT_3}^X &=& -\ca \sodt
 - \sa \cz \codt ,
 \nonumber \\
{\calT_3}^Y &=& \ca \cz \codt
 - \sa \sodt ,
 \nonumber \\
{\calT_3}^Z &=& \sz \codt .
\labeleea{XformComp} 
In these equations,
the abbreviations 
$s_x\equiv \sin x$, $c_x\equiv \cos x$
and $\Om T \equiv \Om_\oplus T$
are used.

\begin{widetext} 

\vskip 10pt

\section{Sun-frame coefficients} 
\citelabel{DefineTildes} 
The Sun-frame tilde coefficients are defined as follows: 
\bea
\tilde{b}_J &=& b_J 
- \half \ve_{JKL}H_{KL} 
- m(d_{JT} - \half \ve_{JKL}g_{KLT}) ,
\qquad
\tilde{b}_T = b_T+mg_{XYZ} ,
\nonumber \\
\tilde{g}_T &=& b_T-m(g_{XYZ}-g_{YZX}-g_{ZXY}) ,
\nonumber \\
\tilde{H}_{XT} &=& H_{XT}+m(d_{ZY}-g_{XTT}-g_{XYY}) ,
\qquad
\tilde{H}_{YT} = H_{YT}+m(d_{XZ}-g_{YTT}-g_{YZZ}) ,
\nonumber \\
\tilde{H}_{ZT} &=& H_{ZT}+m(d_{YX}-g_{ZTT}-g_{ZXX}) ,
\nonumber \\
\tilde{d}_\pm &=& m(d_{XX} \pm d_{YY}) ,
\qquad
\tilde{d}_Q = m(d_{XX}+d_{YY}-2d_{ZZ}-g_{YZX}-g_{ZXY}+2g_{XYZ}) ,
\nonumber \\
\tilde{d}_J &=& m(d_{TJ} + \half d_{JT}) - \frac{1}{4} \ve_{JKL} H_{KL} ,
\qquad
\tilde{d}_{YZ} = m(d_{YZ}+d_{ZY}-g_{XYY} + g_{XZZ}) ,
\nonumber \\
\tilde{d}_{ZX} &=& m(d_{ZX}+d_{XZ}-g_{YZZ} + g_{YXX}) ,
\qquad
\tilde{d}_{XY} = m(d_{XY}+d_{YX}-g_{ZXX} + g_{ZYY}) ,
\nonumber \\
\tilde{g}_c &=& m(g_{XYZ}-g_{ZXY}) ,
\qquad
\tilde{g}_- = m(g_{XTX}-g_{YTY}) ,
\qquad
\tilde{g}_Q = m(g_{XTX}+g_{YTY}-2g_{ZTZ}) ,
\nonumber \\
\tilde{g}_{TJ} &=& m\abs{\ve_{JKL}}g_{KTL} ,
\qquad
\tilde{g}_{DJ} = -b_J+m\ve_{JKL}(g_{KTL}+\half g_{KLT}) ,
\nonumber \\
\tilde{g}_{JK} &=& m(g_{JTT}+g_{JKK})
\quad ({\rm no\ K\ sum},~~J\not= K) ,
\qquad
\tilde{c}_Q = m(c_{XX}+c_{YY}-2c_{ZZ}) ,
\nonumber \\
\tilde{c}_- &=& m(c_{XX}-c_{YY}) ,
\qquad
\tilde{c}_J = m \abs{\ve_{JKL}}c_{KL} , 
\qquad
\tilde{c}_{TJ} = m(c_{TJ}+c_{JT}) .
\labeleea{ListTildes}
Indices $J,K,L$ run over Sun-frame spatial coordinates $X,Y,Z$.
The usual summation convention holds
except where indicated.
The totally antisymmetric tensor $\ve_{JKL}$ 
is defined with $\ve_{XYZ}=+1$. 
Note that 
$\tc_X$, $\tc_Y$, $\tc_Z$,
$\tg_{TX}$, $\tg_{TY}$, and $\tg_{TZ}$
were denoted 
$\tc_{Q,Y}$, $\tc_{Q,X}$, $\tc_{XY}$, 
$\tg_{Q,Y}$, $\tg_{Q,X}$, and $\tg_{XY}$, 
respectively, 
in some previous works. 

\section{Explicit clock-frame coefficients} 
\citelabel{ExplicitTildes} 

This appendix provides the
explicit expressions for the clock-frame tilde coefficients
in terms of the Sun-frame tilde coefficients.
For simplicity, 
we write $\Om T$ for the combination $\Om_\oplus T$
and use the abbreviations $s_x:=\sin x$ and $c_x:=\cos x$
for all trigonometric dependences other than the
relatively rapid $\om_s$ oscillations.

The results are as follows:
\bea 
\tb_3 &=& 
\codt \Big\{ \Big[
 \tb_X(-\sa\cz) + \tb_Y(\ca\cz) + \tb_Z(\sz) 
 \Big]
 \nonumber \\ 
&& +\be_\oplus\Big[ 
 \td_-( -\half\ca\cz\ce\cto + \half\sa\cz\sto ) 
 +\td_+( 2\ca\cz\ce\cto + 2\sz\se\cto + 2\sa\cz\sto ) 
 \nonumber \\ 
&& \phb +\tb_T( -\half\ca\cz\ce\cto + \half\sa\cz\sto ) 
 +\td_Q (-\half\ca\cz\ce\cto-\sz\se\cto-\half\sa\cz\sto) 
 \nonumber \\ 
&& \phb +\td_{XY}(-\sa\cz\ce\cto) + \td_{YZ}(\ca\cz\se\cto) 
 +\td_{ZX}(-\sz\sto) 
 +\tg_c(\ca\cz\ce\cto - \sa\cz\sto) 
 \nonumber \\ 
&& \phb +\tg_T(-\half\ca\cz\ce\cto-\sz\se\cto-\frac{3}{2}\sa\cz\sto) 
 +\tH_{XT}(\sz\ce\cto-\ca\cz\se\cto) 
 \nonumber \\ 
&& \phb +\tH_{YT}(-\sa\cz\se\cto+\sz\sto)
 +\tH_{ZT}(\sa\cz\ce\cto-\ca\cz\sto) 
 \Big] \Big\} 
 \nonumber \\ 
&+& \sodt \Big\{ \Big[
 \tb_X(-\ca) + \tb_Y(-\sa) 
 \Big] 
 \nonumber \\ 
&& +\be_\oplus\Big[ 
 \td_-(\half\sa\ce\cto+\half\ca\sto) 
 +\td_+(-2\sa\ce\cto+2\ca\sto) 
 +\tb_T(\half\sa\ce\cto+\half\ca\sto) 
 \nonumber \\ 
&& \phb +\td_Q(\half\sa\ce\cto-\half\ca\sto) 
 +\td_{XY}(-\ca\ce\cto) + \td_{YZ}(-\sa\se\cto) 
 +\tg_c(-\sa\ce\cto-\ca\sto) 
 \nonumber \\ 
&& \phb +\tg_T(\half\sa\ce\cto-\frac{3}{2}\ca\sto) 
 +\tH_{XT}(\sa\se\cto) 
 +\tH_{YT}(-\ca\se\cto) 
 +\tH_{ZT}(\ca\ce\cto+\sa\sto) 
 \Big] \Big\} 
 \nonumber \\ 
&+& \ctodt \Big\{ 
 \be_s\Big[ 
 \td_-(\frac{3}{8}\cta+\frac{1}{8}\cta\ctz) 
 +\tb_T(\frac{3}{8}\cta+\frac{1}{8}\cta\ctz) 
 +\td_Q(\frac{1}{8}-\frac{1}{8}\ctz) 
 +\td_{XY}(\frac{3}{8}\sta+\frac{1}{8}\sta\ctz) 
 \nonumber \\ 
&& \phb + \td_{YZ}(-\frac{1}{4}\ca\stz) 
 +\td_{ZX}(\frac{1}{4}\sa\stz) 
 +\tg_c(-\frac{3}{4}\cta-\frac{1}{4}\cta\ctz) 
 +\tg_T(-\frac{3}{8}\cta-\frac{1}{8}\cta\ctz) 
 \Big] \Big\} 
 \nonumber \\ 
&+& \stodt \Big\{ 
 \be_s\Big[ 
 \td_-(-\half\sta\cz) 
 +\tb_T(-\half\sta\cz) 
 +\td_{XY}(\half\cta\cz) 
 + \td_{YZ}(\half\sa\sz) 
 +\td_{ZX}(\half\ca\sz) 
 \nonumber \\ 
&& \phb +\tg_c(\sta\cz) 
 +\tg_T(\half\sta\cz) 
 \Big] \Big\} 
 \nonumber \\ 
&+& \Big\{ 
 \be_s\Big[ 
 \td_-(-\frac{1}{8}\cta+\frac{1}{8}\cta\ctz) 
 + \td_+(-2) 
 +\tb_T(-\frac{1}{8}\cta+\frac{1}{8}\cta\ctz) 
 +\td_Q(\frac{5}{8}-\frac{1}{8}\ctz) 
 +\td_{XY}(-\frac{1}{8}\sta+\frac{1}{8}\sta\ctz) 
 \nonumber \\ 
&& \phb + \td_{YZ}(-\frac{1}{4}\ca\stz) 
 +\td_{ZX}(\frac{1}{4}\sa\stz) 
 +\tg_c(\frac{1}{4}\cta-\frac{1}{4}\cta\ctz) 
 +\tg_T(1+\frac{1}{8}\cta-\frac{1}{8}\cta\ctz) 
 \Big] \Big\} ,
\labeleea{Explicitb3}  
\bea 
\td_3 &=& 
\codt \Big\{ \Big[ 
 \td_X(-\sa\cz) + \td_Y(\ca\cz) + \td_Z(\sz) 
 \Big] 
 \nonumber \\ 
&& +\be_\oplus\Big[ 
 \td_-( \frac{3}{4}\ca\cz\ce\cto-\frac{3}{4}\sa\cz\sto ) 
 +\td_+( -3\ca\cz\ce\cto-3\sz\se\cto-3\sa\cz\sto ) 
 \nonumber \\ 
&& \phb +\tb_T( -\frac{3}{4}\ca\cz\ce\cto-\frac{3}{2}\sz\se\cto-\frac{3}{4}\sa\cz\sto ) 
 +\td_Q ( \frac{3}{4}\ca\cz\ce\cto+\frac{3}{2}\sz\se\cto+\frac{3}{4}\sa\cz\sto ) 
 \nonumber \\ 
&& \phb +\td_{XY}( \frac{1}{2}\sa\cz\ce\cto+\ca\cz\sto ) 
 +\td_{YZ}( -\sz\ce\cto-\frac{1}{2}\ca\cz\se\cto ) 
 +\td_{ZX}( \sa\cz\se\cto+\frac{1}{2}\sz\sto ) 
 \nonumber \\ 
&& \phb +\tg_T( \frac{3}{4}\ca\cz\ce\cto+\frac{3}{2}\sz\se\cto+\frac{3}{4}\sa\cz\sto ) 
 +\tg_{XY}( -\frac{1}{2}\sz\ce\cto-\ca\cz\se\cto ) 
 \nonumber \\ 
&& \phb +\tg_{XZ}( \sz\ce\cto+\frac{1}{2}\ca\cz\se\cto )
 +\tg_{YX}( -\sa\cz\se\cto-\frac{1}{2}\sz\sto ) 
 +\tg_{YZ}( \frac{1}{2}\sa\cz\se\cto+\sz\sto ) 
 \nonumber \\ 
&& \phb +\tg_{ZX}( \sa\cz\ce\cto+\frac{1}{2}\ca\cz\sto ) 
 +\tg_{ZY}( -\frac{1}{2}\sa\cz\ce\cto-\ca\cz\sto ) 
 \nonumber \\ 
&& \phb +\tH_{XT}( \frac{1}{2}\sz\ce\cto-\frac{1}{2}\ca\cz\se\cto ) 
 +\tH_{YT}( -\frac{1}{2}\sa\cz\se\cto+\frac{1}{2}\sz\sto )
 +\tH_{ZT}( \frac{1}{2}\sa\cz\ce\cto-\frac{1}{2}\ca\cz\sto ) 
 \Big] \Big\} 
 \nonumber \\ 
&+& \sodt \Big\{ \Big[ 
 \td_X( -\ca ) + \td_Y( -\sa ) 
 \Big] 
 \nonumber \\ 
&& +\be_\oplus\Big[ 
 \td_-( -\frac{3}{4}\sa\ce\cto-\frac{3}{4}\ca\sto ) 
 +\td_+( 3\sa\ce\cto-3\ca\sto ) 
 +\tb_T( \frac{3}{4}\sa\ce\cto-\frac{3}{4}\ca\sto ) 
 \nonumber \\ 
&& \phb +\td_Q ( -\frac{3}{4}\sa\ce\cto+\frac{3}{4}\ca\sto ) 
 +\td_{XY}( \half\ca\ce\cto-\sa\sto ) 
 +\td_{YZ}( \half\sa\se\cto ) 
 +\td_{ZX}( \ca\se\cto ) 
 \nonumber \\ 
&& \phb +\tg_T( -\frac{3}{4}\sa\ce\cto+\frac{3}{4}\ca\sto ) 
 +\tg_{XY}( \sa\se\cto ) 
 +\tg_{XZ}( -\half\sa\se\cto )
 +\tg_{YX}( -\ca\se\cto ) 
 \nonumber \\ 
&& \phb +\tg_{YZ}( \half\ca\se\cto ) 
 +\tg_{ZX}( \ca\ce\cto-\half\sa\sto ) 
 +\tg_{ZY}( -\half\ca\ce\cto+\sa\sto ) 
 \nonumber \\ 
&& \phb +\tH_{XT}( \half\sa\se\cto ) 
 +\tH_{YT}( -\half\ca\se\cto )
 +\tH_{ZT}( \half\ca\ce\cto+\half\sa\sto ) 
 \Big] \Big\} 
 \nonumber \\ 
&+& \ctodt \Big\{ 
 \be_s\Big[ 
 \td_-( -\frac{9}{16}\cta-\frac{3}{16}\cta\ctz ) 
 +\tb_T( \frac{3}{16}-\frac{3}{16}\ctz ) 
 +\td_Q ( -\frac{3}{16}+\frac{3}{16}\ctz ) 
 \nonumber \\ 
&& \phb +\td_{XY}( -\frac{9}{16}\sta-\frac{3}{16}\sta\ctz ) 
 +\td_{YZ}( \frac{3}{8}\ca\stz ) 
 +\td_{ZX}( -\frac{3}{8}\sa\stz ) 
 +\tg_T( -\frac{3}{16}+\frac{3}{16}\ctz ) 
 \nonumber \\ 
&& \phb +\tg_{XY}( \frac{3}{8}\ca\stz ) 
 +\tg_{XZ}( -\frac{3}{8}\ca\stz )
 +\tg_{YX}( \frac{3}{8}\sa\stz ) 
 +\tg_{YZ}( -\frac{3}{8}\sa\stz ) 
 \nonumber \\ 
&& \phb +\tg_{ZX}( -\frac{9}{16}\sta-\frac{3}{16}\sta\ctz ) 
 +\tg_{ZY}( \frac{9}{16}\sta+\frac{3}{16}\sta\ctz ) 
 \Big] \Big\} 
 \nonumber \\ 
&+& \stodt \Big\{ 
 \be_s\Big[ 
 \td_-( \frac{3}{4}\sta\cz ) 
 +\td_{XY}( -\frac{3}{4}\cta\cz ) 
 +\td_{YZ}( -\frac{3}{4}\sa\sz ) 
 +\td_{ZX}( -\frac{3}{4}\ca\sz ) 
 \nonumber \\ 
&& \phb +\tg_{XY}( -\frac{3}{4}\sa\sz ) 
 +\tg_{XZ}( \frac{3}{4}\sa\sz )
 +\tg_{YX}( \frac{3}{4}\ca\sz ) 
 +\tg_{YZ}( -\frac{3}{4}\ca\sz ) 
 \nonumber \\ 
&& \phb +\tg_{ZX}( -\frac{3}{4}\cta\cz ) 
 +\tg_{ZY}( \frac{3}{4}\cta\cz ) 
 \Big] \Big\} 
 \nonumber \\ 
&+& \Big\{ 
 \be_s\Big[ 
 \td_-( \frac{3}{16}\cta-\frac{3}{16}\cta\ctz ) 
 +\td_+( 3 ) 
 +\tb_T( \frac{15}{16}-\frac{3}{16}\ctz ) 
 +\td_Q ( -\frac{15}{16}+\frac{3}{16}\ctz ) 
 +\td_{XY}( \frac{3}{16}\sta-\frac{3}{16}\sta\ctz ) 
 \nonumber \\ 
&& \phb +\td_{YZ}( \frac{3}{8}\ca\stz ) 
 +\td_{ZX}( -\frac{3}{8}\sa\stz ) 
 +\tg_T( -\frac{15}{16}+\frac{3}{16}\ctz ) 
 +\tg_{XY}( \frac{3}{8}\ca\stz ) 
 +\tg_{XZ}( -\frac{3}{8}\ca\stz )
 \nonumber \\ 
&& \phb +\tg_{YX}( \frac{3}{8}\sa\stz ) 
 +\tg_{YZ}( -\frac{3}{8}\sa\stz ) 
 +\tg_{ZX}( \frac{3}{16}\sta-\frac{3}{16}\sta\ctz ) 
 +\tg_{ZY}( -\frac{3}{16}\sta+\frac{3}{16}\sta\ctz ) 
 \Big] \Big\} ,
\labeleea{Explicitd3}  
\bea 
\tg_d &=& 
\codt \Big\{ \Big[ 
 \tg_{DX}( -\sa\cz ) + \tg_{DY}( \ca\cz ) + \tg_{DZ}( \sz ) 
 \Big] 
 \nonumber \\ 
&& +\be_\oplus\Big[ 
  \tb_T( 2\sa\cz\sto ) 
 +\tg_c( 2\ca\cz\ce\cto-2\sa\cz\sto ) 
 +\tg_T( \ca\cz\ce\cto+\sz\se\cto-\sa\cz\sto ) 
 \nonumber \\ 
&& \phb +\tg_{XY}( -2\sz\ce\cto ) 
 +\tg_{XZ}( 2\ca\cz\se\cto )
 +\tg_{YX}( -2\sz\sto ) 
 +\tg_{YZ}( 2\sa\cz\se\cto ) 
 \nonumber \\ 
&& \phb +\tg_{ZX}( 2\ca\cz\sto ) 
 +\tg_{ZY}( -2\sa\cz\ce\cto ) 
 \Big] \Big\} 
 \nonumber \\ 
&+& \sodt \Big\{ \Big[ 
 \tg_{DX}( -\ca ) + \tg_{DY}( -\sa ) 
 \Big] 
 \nonumber \\ 
&& +\be_\oplus\Big[ \tb_T( 2\ca\sto ) 
 +\tg_c( -2\sa\ce\cto-2\ca\sto ) 
 +\tg_T( -\sa\ce\cto-\ca\sto ) 
 \nonumber \\ 
&& \phb +\tg_{XZ}( -2\sa\se\cto )
 +\tg_{YZ}( 2\ca\se\cto ) 
 +\tg_{ZX}( -2\sa\sto ) 
 +\tg_{ZY}( -2\ca\ce\cto ) 
 \Big] \Big\} 
 \nonumber \\ 
&+& \ctodt \Big\{ 
 \be_s\Big[ 
  \tb_T( \frac{1}{4}+\frac{3}{4}\cta-\frac{1}{4}\ctz+\frac{1}{4}\cta\ctz ) 
 +\tg_c( -\frac{3}{2}\cta-\half\cta\ctz ) 
 \nonumber \\ 
&& \phb +\tg_T( -\frac{1}{4}-\frac{3}{4}\cta+\frac{1}{4}\ctz-\frac{1}{4}\cta\ctz ) 
 +\tg_{XY}( \half\ca\stz ) 
 +\tg_{XZ}( -\half\ca\stz )
 +\tg_{YX}( \half\sa\stz ) 
 +\tg_{YZ}( -\half\sa\stz ) 
 \nonumber \\ 
&& \phb +\tg_{ZX}( -\frac{3}{4}\sta-\frac{1}{4}\sta\ctz ) 
 +\tg_{ZY}( \frac{3}{4}\sta+\frac{1}{4}\sta\ctz ) 
 \Big] \Big\} 
 \nonumber \\ 
&+& \stodt \Big\{ 
 \be_s\Big[ 
  \tb_T( -\sta\cz ) 
 +\tg_c( 2\sta\cz ) 
 +\tg_T( \sta\cz ) 
 +\tg_{XY}( -\sa\sz ) 
 +\tg_{XZ}( \sa\sz )
 \nonumber \\ 
&& \phb +\tg_{YX}( \ca\sz ) 
 +\tg_{YZ}( -\ca\sz ) 
 +\tg_{ZX}( -\cta\cz ) 
 +\tg_{ZY}( \cta\cz ) 
 \Big] \Big\} 
 \nonumber \\ 
&+& \Big\{ 
 \be_s\Big[ 
  \tb_T( -\frac{3}{4}-\frac{1}{4}\cta-\frac{1}{4}\ctz+\frac{1}{4}\cta\ctz ) 
 +\tg_c( \half\cta-\half\cta\ctz ) 
 +\tg_T( -\frac{1}{4}+\frac{1}{4}\cta+\frac{1}{4}\ctz-\frac{1}{4}\cta\ctz ) 
 \nonumber \\ 
&& \phb +\tg_{XY}( \half\ca\stz ) 
 +\tg_{XZ}( -\half\ca\stz )
 +\tg_{YX}( \half\sa\stz ) 
 +\tg_{YZ}( -\half\sa\stz ) 
 \nonumber \\ 
&& \phb +\tg_{ZX}( \frac{1}{4}\sta-\frac{1}{4}\sta\ctz ) 
 +\tg_{ZY}( -\frac{1}{4}\sta+\frac{1}{4}\sta\ctz ) 
 \Big] \Big\} ,
\labeleea{Explicitgd}  
\bea 
\tc_q &=& 
\codt \Big\{ 
 \be_s\Big[ 
  \tc_{TX}( 2\sa\cz ) 
 +\tc_{TY}( -2\ca\cz ) 
 +\tc_{TZ}( -2\sz ) 
 \Big] \Big\} 
 \nonumber \\ 
&+& \sodt \Big\{ 
 \be_s\Big[ 
  \tc_{TX}( 2\ca ) 
 +\tc_{TY}( 2\sa ) 
 \Big] \Big\} 
 \nonumber \\ 
&+& \ctodt \Big\{ \Big[ 
  \tc_-( \frac{9}{8}\cta+\frac{3}{8}\cta\ctz ) 
 +\tc_Q( \frac{3}{8}-\frac{3}{8}\ctz ) 
 +\tc_X( -\frac{3}{4}\ca\stz ) 
 +\tc_Y( \frac{3}{4}\sa\stz ) 
 +\tc_Z( \frac{9}{8}\sta+\frac{3}{8}\sta\ctz ) 
 \Big] 
 \nonumber \\ 
&& +\be_\oplus\Big[ 
  \tc_{TX}( -\frac{9}{8}\sta\ce\cto-\frac{3}{8}\sta\ctz\ce\cto-\frac{3}{4}\sa\stz\se\cto
  +\frac{3}{8}\sto+\frac{9}{8}\cta\sto-\frac{3}{8}\ctz\sto+\frac{3}{8}\cta\ctz\sto ) 
 \nonumber \\ 
&& \phb +\tc_{TY}( -\frac{3}{8}\ce\cto+\frac{9}{8}\cta\ce\cto+\frac{3}{8}\ctz\ce\cto 
 +\frac{3}{8}\cta\ctz\ce\cto 
 +\frac{3}{4}\ca\stz\se\cto+\frac{9}{8}\sta\sto+\frac{3}{8}\sta\ctz\sto ) 
 \nonumber \\ 
&& \phb +\tc_{TZ}( \frac{3}{4}\ca\stz\ce\cto+\frac{3}{4}\se\cto
 -\frac{3}{4}\ctz\se\cto+\frac{3}{4}\sa\stz\sto ) 
 \Big] \Big\} 
 \nonumber \\ 
&+& \stodt \Big\{ \Big[ 
  \tc_-( -\frac{3}{2}\sta\cz ) 
 +\tc_X( \frac{3}{2}\sa\sz ) 
 +\tc_Y( \frac{3}{2}\ca\sz ) 
 +\tc_Z( \frac{3}{2}\cta\cz ) 
 \Big] 
 \nonumber \\ 
&& +\be_\oplus\Big[ 
  \tc_{TX}( -\frac{3}{2}\cta\cz\ce\cto-\frac{3}{2}\ca\sz\se\cto-\frac{3}{2}\sta\cz\sto ) 
 \nonumber \\ 
&& \phb 
 +\tc_{TY}( -\frac{3}{2}\sta\cz\ce\cto-\frac{3}{2}\sa\sz\se\cto+\frac{3}{2}\cta\cz\sto ) 
 +\tc_{TZ}( -\frac{3}{2}\sa\sz\ce\cto+\frac{3}{2}\ca\sz\sto ) 
 \Big] \Big\} 
 \nonumber \\ 
&+& \Big\{ \Big[
  \tc_-( -\frac{3}{8}\cta+\frac{3}{8}\cta\ctz ) 
 +\tc_Q( -\frac{1}{8}-\frac{3}{8}\ctz ) 
 +\tc_X( -\frac{3}{4}\ca\stz ) 
 +\tc_Y( \frac{3}{4}\sa\stz ) 
 +\tc_Z( -\frac{3}{8}\sta+\frac{3}{8}\sta\ctz ) 
 \Big] 
 \nonumber \\ 
&& +\be_\oplus\Big[ 
  \tc_{TX}( \frac{3}{8}\sta\ce\cto-\frac{3}{8}\sta\ctz\ce\cto-\frac{3}{4}\sa\stz\se\cto
 -\frac{1}{8}\sto-\frac{3}{8}\cta\sto-\frac{3}{8}\ctz\sto+\frac{3}{8}\cta\ctz\sto ) 
 \nonumber \\ 
&& \phb +\tc_{TY}( \frac{1}{8}\ce\cto-\frac{3}{8}\cta\ce\cto+\frac{3}{8}\ctz\ce\cto 
  +\frac{3}{8}\cta\ctz\ce\cto
 +\frac{3}{4}\ca\stz\se\cto-\frac{3}{8}\sta\sto+\frac{3}{8}\sta\ctz\sto ) 
 \nonumber \\ 
&& \phb +\tc_{TZ}( \frac{3}{4}\ca\stz\ce\cto-\frac{1}{4}\se\cto-\frac{3}{4}\ctz\se\cto 
  +\frac{3}{4}\sa\stz\sto ) 
 \Big] \Big\} ,
\labeleea{Explicitcq}  
\bea 
\tg_q &=& 
\codt \Big\{ 
 \be_s\Big[ 
  \tg_{XY}( \sa\cz ) 
 +\tg_{XZ}( \sa\cz ) 
 +\tg_{YX}( -\ca\cz ) 
 +\tg_{YZ}( -\ca\cz ) 
 +\tg_{ZX}( -\sz ) 
 +\tg_{ZY}( -\sz ) 
 \Big] 
 \nonumber \\ 
&+& \sodt \Big\{ 
 \be_s\Big[ 
  \tg_{XY}( \ca ) 
 +\tg_{XZ}( \ca ) 
 +\tg_{YX}( \sa ) 
 +\tg_{YZ}( \sa ) 
 \Big] 
 \nonumber \\ 
&+& \ctodt \Big\{ \Big[ 
  \tg_-( \frac{9}{8}\cta+\frac{3}{8}\cta\ctz ) 
 +\tg_Q( \frac{3}{8}-\frac{3}{8}\ctz ) 
 +\tg_{TX}( -\frac{3}{4}\ca\stz ) 
 +\tg_{TY}( \frac{3}{4}\sa\stz ) 
 +\tg_{TZ}( \frac{9}{8}\sta+\frac{3}{8}\sta\ctz ) 
 \Big] 
 \nonumber \\ 
&& +\be_\oplus\Big[ 
  \tb_T( -\frac{3}{4}\sa\stz\ce\cto+\frac{9}{8}\sta\se\cto+\frac{3}{8}\sta\ctz\se\cto ) 
 \nonumber \\ 
&& \phb +\tg_c( \frac{3}{4}\sa\stz\ce\cto-\frac{9}{4}\sta\se\cto-\frac{3}{4}\sta\ctz\se\cto 
  +\frac{3}{4}\ca\stz\sto ) 
 \nonumber \\ 
&& \phb +\tg_T( \frac{3}{4}\sa\stz\ce\cto-\frac{9}{8}\sta\se\cto-\frac{3}{8}\sta\ctz\se\cto ) 
 \nonumber \\ 
&& \phb +\tg_{XY}( -\frac{9}{8}\sta\ce\cto-\frac{3}{8}\sta\ctz\ce\cto-\frac{3}{8}\sto 
 +\frac{9}{8}\cta\sto+\frac{3}{8}\ctz\sto+\frac{3}{8}\cta\ctz\sto )
 \nonumber \\
&& \phb +\tg_{XZ}( -\frac{3}{4}\sa\stz\se\cto+\frac{3}{4}\sto-\frac{3}{4}\ctz\sto ) 
 \nonumber \\ 
&& \phb +\tg_{YX}( \frac{3}{8}\ce\cto+\frac{9}{8}\cta\ce\cto-\frac{3}{8}\ctz\ce\cto 
 +\frac{3}{8}\cta\ctz\ce\cto+\frac{9}{8}\sta\sto+\frac{3}{8}\sta\ctz\sto ) 
 \nonumber \\ 
&& \phb +\tg_{YZ}( -\frac{3}{4}\ce\cto+\frac{3}{4}\ctz\ce\cto+\frac{3}{4}\ca\stz\se\cto ) 
 \nonumber \\ 
&& \phb +\tg_{ZX}( \frac{3}{8}\se\cto+\frac{9}{8}\cta\se\cto-\frac{3}{8}\ctz\se\cto 
 +\frac{3}{8}\cta\ctz\se\cto+\frac{3}{4}\sa\stz\sto ) 
 \nonumber \\ 
&& \phb +\tg_{ZY}( \frac{3}{4}\ca\stz\ce\cto+\frac{3}{8}\se\cto-\frac{9}{8}\cta\se\cto
 -\frac{3}{8}\ctz\se\cto-\frac{3}{8}\cta\ctz\se\cto ) 
 \Big] \Big\} 
 \nonumber \\ 
&+& \stodt \Big\{ \Big[ 
  \tg_-( -\frac{3}{2}\sta\cz ) 
 +\tg_{TX}( \frac{3}{2}\sa\sz ) 
 +\tg_{TY}( \frac{3}{2}\ca\sz ) 
 +\tg_{TZ}( \frac{3}{2}\cta\cz ) 
 \Big] 
 \nonumber \\ 
&& +\be_\oplus\Big[ 
  \tb_T( -\frac{3}{2}\ca\sz\ce\cto+\frac{3}{2}\cta\cz\se\cto ) 
 +\tg_c( \frac{3}{2}\ca\sz\ce\cto-3\cta\cz\se\cto-\frac{3}{2}\sa\sz\sto ) 
 \nonumber \\ 
&& \phb +\tg_T( \frac{3}{2}\ca\sz\ce\cto-\frac{3}{2}\cta\cz\se\cto ) 
 \nonumber \\ 
&& \phb +\tg_{XY}( -\frac{3}{2}\cta\cz\ce\cto-\frac{3}{2}\sta\cz\sto ) 
 +\tg_{XZ}( -\frac{3}{2}\ca\sz\se\cto ) 
 +\tg_{YX}( -\frac{3}{2}\sta\cz\ce\cto+\frac{3}{2}\cta\cz\sto ) 
 \nonumber \\ 
&& \phb +\tg_{YZ}( -\frac{3}{2}\sa\sz\se\cto ) 
 +\tg_{ZX}( -\frac{3}{2}\sta\cz\se\cto+\frac{3}{2}\ca\sz\sto ) 
 +\tg_{ZY}( -\frac{3}{2}\sa\sz\ce\cto+\frac{3}{2}\sta\cz\se\cto ) 
 \Big] \Big\} 
 \nonumber \\ 
&+& \Big\{ \Big[
  \tg_-( -\frac{3}{8}\cta+\frac{3}{8}\cta\ctz ) 
 +\tg_Q( -\frac{1}{8}-\frac{3}{8}\ctz ) 
 +\tg_{TX}( -\frac{3}{4}\ca\stz ) 
 +\tg_{TY}( \frac{3}{4}\sa\stz ) 
 +\tg_{TZ}( -\frac{3}{8}\sta+\frac{3}{8}\sta\ctz ) 
 \Big] 
 \nonumber \\ 
&& +\be_\oplus\Big[ 
  \tb_T( -\frac{3}{4}\sa\stz\ce\cto-\frac{3}{8}\sta\se\cto+\frac{3}{8}\sta\ctz\se\cto ) 
 \nonumber \\ 
&& \phb +\tg_c( \frac{3}{4}\sa\stz\ce\cto+\frac{3}{4}\sta\se\cto-\frac{3}{4}\sta\ctz\se\cto 
  +\frac{3}{4}\ca\stz\sto ) 
 \nonumber \\ 
&& \phb +\tg_T( \frac{3}{4}\sa\stz\ce\cto+\frac{3}{8}\sta\se\cto-\frac{3}{8}\sta\ctz\se\cto ) 
 \nonumber \\ 
&& \phb +\tg_{XY}( \frac{3}{8}\sta\ce\cto-\frac{3}{8}\sta\ctz\ce\cto+\frac{1}{8}\sto 
 -\frac{3}{8}\cta\sto+\frac{3}{8}\ctz\sto+\frac{3}{8}\cta\ctz\sto ) 
 \nonumber \\ 
&& \phb +\tg_{XZ}( -\frac{3}{4}\sa\stz\se\cto-\frac{1}{4}\sto-\frac{3}{4}\ctz\sto ) 
 \nonumber \\ 
&& \phb +\tg_{YX}( -\frac{1}{8}\ce\cto-\frac{3}{8}\cta\ce\cto-\frac{3}{8}\ctz\ce\cto 
 +\frac{3}{8}\cta\ctz\ce\cto-\frac{3}{8}\sta\sto+\frac{3}{8}\sta\ctz\sto ) 
 \nonumber \\ 
&& \phb +\tg_{YZ}( \frac{1}{4}\ce\cto+\frac{3}{4}\ctz\ce\cto+\frac{3}{4}\ca\stz\se\cto ) 
 \nonumber \\ 
&& \phb +\tg_{ZX}( -\frac{1}{8}\se\cto-\frac{3}{8}\cta\se\cto-\frac{3}{8}\ctz\se\cto 
 +\frac{3}{8}\cta\ctz\se\cto+\frac{3}{4}\sa\stz\sto ) 
 \nonumber \\ 
&& \phb +\tg_{ZY}( \frac{3}{4}\ca\stz\ce\cto-\frac{1}{8}\se\cto+\frac{3}{8}\cta\se\cto 
 -\frac{3}{8}\ctz\se\cto-\frac{3}{8}\cta\ctz\se\cto ) 
 \Big] \Big\} .
\labeleea{Explicitgq}  

\end{widetext}


\begin{references}
\bibitem{fn1}
This assumption fails for some experimental arrangements,
such as an ion clock operating on a field-insensitive line,
but the properties of $\som$ in these cases 
can be treated by other methods.
\bibitem{fn2}
In certain experiments,
the rotation of the clock within the laboratory
(for example, on a turntable in an Earth-based experiment)
must also be incorporated.
This can be implemented within our framework
but is irrelevant for present purposes.
The procedure is described in the context of
classic relativity tests with light in Ref.\ \cite{km}.
\bibitem{fn3}
Some types of Earth-based experiments 
can be studied as a limiting case of the 
following analysis,
in particular ones in which the clock quantization axis
is fixed and directed to the east.
The relevant limits are:
$r_s\to r_L\approx (6380\ {\rm km})\cos\la$,
which gives the effective radius of the Earth at the latitude
$\la$ of the laboratory $L$;
$\om_s\to\om_L= 2\pi/$(1 sidereal day);
$\be_s\to\be_L=r_L \om_L$;
$\ze\to 0$; $\al\to 0$;
and $T_s \to T_\oplus$.
The choice of zero $T_0$ must then be interpreted
as in Appendix C 1 of Ref.\ \cite{km}.
\end{references}
\end{document}